\documentclass[preprint,10pt,fleqn,authoryear]{elsarticle}

\usepackage[section]{placeins}
\usepackage{svg}
\usepackage{amsfonts}
\usepackage{amssymb}
\usepackage[top=2cm,bottom=2.5cm,right=1.5cm,left=1.5cm]{geometry}
\usepackage{amsthm}
\usepackage{amsmath}
\usepackage{graphicx}
\usepackage{caption}
\usepackage[colorlinks=true, allcolors=blue]{hyperref}
\usepackage{bm}

\newcommand{\comment}{}
\def \d{{\textrm d}}

\begin{document}

\captionsetup[figure]{labelfont={bf},labelformat={default},labelsep=period,name={Fig. }}

\begin{frontmatter}



\title{An exact dimension-reduced dynamic theory for developable surfaces and curve-fold origami}


\author[inst1]{Zhixuan Wen}

\affiliation[inst1]{organization={State Key Laboratory for Turbulence and Complex Systems, School of Mechanics and Engineering Science, Peking University},
            city={Beijing},
            postcode={100871}, 
            country={China}}

\author[inst3]{Sheng Mao}
\author[inst1,inst2]{Huiling Duan}
\author[inst3]{Fan Feng \corref{cor1}}

\affiliation[inst2]{organization={HEDPS, CAPT and IFSA, Collaborative Innovation Center of MoE, Peking University},
            city={Beijing},
            postcode={100871}, 
            country={China}}

\affiliation[inst3]{organization={School of Mechanics and Engineering Science, Peking University},
            city={Beijing},
            postcode={100871}, 
            country={China}}
\cortext[cor1]{\noindent Corresponding author \newline  \indent E-mail address: fanfeng@pku.edu.cn (Fan Feng)}  
    
\begin{abstract}
Curve-fold origami, composed of developable panels joined along a curved crease, exhibits rich dynamic behaviors relevant to metamaterials and soft robotic systems.  Despite multiple approximated models ({shell model, Kirchhoff rod model, etc.}), a comprehensive and exact dynamical theory for curve-fold origami remains absent, limiting the precise prediction of its dynamics, especially for those with wide panels. 
In this work, we develop an exact dimension-reduced theory that focuses on the dynamics of curve-fold origami, utilizing the intrinsic one-dimensional nature of developable surfaces. As is well known in differential geometry, a curve on a developable surface determines the local frame and the shape of the developable surface under isometric constraint, making the problem indeed one-dimensional. 
Starting from a single developable surface, we investigate the kinematics and kinetic energy of a moving developable surface. 
By overcoming the difficulty of describing the motion of local frames, we derive the exact velocity field of wide surfaces solely described by the motion of the reference curve, which leads to the kinetic energy of the entire surface.
The elastic energy follows the Wunderlich functional applicable to wide panels. Owing to the one-dimensional feature,  the Lagrangian of the system,
composed of both kinetic and elastic energy, is a functional of the reference curve. Thus, we may variate the Lagrangian and derive a nonlinear dynamical theory for the reference curve, which comprises governing equations similar to the rod model but can precisely describe the motion of developable surfaces. 
The theory is validated consistently in both Lagrangian and Eulerian frameworks and is further extended to curved-fold origami modeled as a coupled bi-rod system.
Utilizing our exact 1D model, we theoretically analyze the dynamical behaviors of various developable structures, revealing that the coupling of curvature and torsion along with the motion of local frames in our theory leads to the accurate modeling of arbitrarily deformed developable surfaces, which are validated by finite element analysis quantitatively. Our exact dimension-reduced dynamical theory captures the motions of developable surfaces and curve-fold origami (with wide panels) accurately at relatively low computational cost (since it is 1D), offering a potential contribution to the design of dynamical developable structures.

\end{abstract}


\end{frontmatter}

\section{Introduction}
\label{sec1}
Origami, a traditional art form, enables complex three-dimensional shape transformations by folding along prescribed crease patterns, and also inspires diverse engineering applications. Over the past centuries, efforts have been made by origamists, mathematicians and physicists on the geometry and statics of straight-fold origami \citep{izmestiev2017classification,tachi2009generalization,demaine2017origamizer,liu2017nonlinear,filipov2017bar}, exploring the quasi-static transformations of origami under geometric constraints. More recently, these efforts have been expanded to the kinematics and dynamics of straight-fold systems, producing models that couple the geometric compatibility with elasticity to describe the motion of dynamic origami structures \citep{zou2024kinematics,herkal2024dynamic}. Such theories have facilitated the development of origami-based metamaterials \citep{doi:10.1126/science.1252876, liu2018topological} and robotic mechanisms \citep{doi:10.1126/scirobotics.aat0938}. Beyond traditional straight folds, recently a novel branch of origami with curved folds, i.e., curve-fold origami, has gained more interest \citep{CurvedCrease_AAG2008, demaine2011curved}. In addition to aesthetic design, curved folds offer unique mechanical capabilities—such as tunable stiffness, energy storage, and large shape change—that have motivated applications in soft robotics \citep{doi:10.1126/scirobotics.aat0938,feng2024geometry}, metamaterials \citep{lee2021compliant, PRLmeta, meta3}, architecture \citep{tachi2011designing, mouthuy2012overcurvature}, and virtual-reality interfaces \citep{VR}.  Besides, its extraordinary dynamic behaviors have enabled the designs of fast-develop robotic wings \citep{doi:10.1126/scirobotics.aaz6262} and acoustic metamaterials \citep{doi:10.1126/sciadv.abe2000}. Despite these advances, we notice that most of the dynamic theories rely on narrow ribbon or rod approximations, and 
an exact dynamical theory for curve-fold origami, in particular for those with wide panels, is still lacking, limiting further applications that require accurate predictions of dynamics at moderate computational cost. Addressing this gap is the primary objective of the present study.

Previous studies on the geometry and statics of curve-fold origami
are
under isometric assumptions, since the paper-like panels are unstretchable. Therefore, the panels will deform as developable surfaces.  For the geometry, building on classical differential geometry, seminal work by David \citet{1674542} and subsequent developments \citep{duncan1982folded,fuchs1999more,kilian2008curved,demaine2015characterization} established a foundational theorem: the configuration of a curved-fold origami or, more generally, a developable surface is completely determined by the geometry of a single deformed curve on the surface—such as the crease in curved-fold origami.  In other words, the distribution of curvature and torsion of the curve uniquely determines the shape of the entire surface. This theorem reveals an essential property of developable surfaces: despite being two-dimensional objects, their admissible deformations under isometric constraints are basically 1D, governed by the geometry of an underlying reference curve.
For the statics, the differential geometry framework enables the modeling of panels as inextensible elastic plates with bending energy proportional to the mean curvature square. Following the geometry,  the bending energy can be expressed as a functional of the curvature $\kappa$ and torsion $\tau$ of the reference curve, with the exact form given by the \citet{Wunderlich1962}  functional.  
Several approximated models have been developed for specific regimes, including the Sadowsky functional for narrow ribbons \citep{sadowsky1930theorie} and Kirchhoff rod formulations for rod-like cross-sections. A comprehensive discussion of the applicability of these models—extensible versus inextensible, rod versus wide or narrow ribbons—is provided by \citet{AUDOLY2021104457}.
In addition to the developable panels, the crease is typically modeled as a torsional spring, and the equilibrium configuration is obtained by minimizing the combined energy of the crease and panels, yielding shape equations governed by the curvature and torsion of the crease. This framework has been employed to analyze the deformation of folded circular strips \citep{dias2012geometric}, multi–curved-fold origami \citep{dias2012thesis, liu_design_2024}, and curved-fold origami with vertices \citep{WEN2024105829}.

In contrast to the well-established geometric and static frameworks for developable surfaces and curved-fold origami, their dynamic behaviors and, in particular, the associated kinetic energy, remain poorly developed.
 Existing studies on dynamics are largely based on the dynamic Kirchhoff rod model for narrow and straight developable ribbons, according to previous works by \citet{Goriely2001}, \citet{DAI2022108511} and \citet{HUANG2024105721}. 
Solving the Kirchhoff equations yields the motion of such structures. These models perform reasonably well for narrow ribbons \citep{doi:10.1073/pnas.2209048120,SHI2025105922}. However, the Kirchhoff rod formulation does not enforce the isometric constraint of developable surfaces, leading to significant errors for highly twisted ribbons \citep{AUDOLY2021104457}, such as Möbius bands \citep{doi:10.1098/rspa.1993.0009,starostin2007shape}.
From the perspective of energy, this limitation arises because the line energy density of Kirchhoff rods scales as $\varepsilon_{crease}\in (A \kappa^2+B\tau^2)$, which becomes inaccurate for configurations with large torsion (i.e., extremely curved case). In contrast, the Sadowsky functional exhibits the scaling $\varepsilon_{crease}\in(\kappa^2+\tau^2)^2/\kappa^2$.
To address this issue, \citet{DIAS201457,dias2015wunderlich} developed a nonlinear rod model for narrow developable surfaces, employing constitutive laws derived from Sadowsky's energy and enforcing the developability/isometric constraint. 
Despite these advances, current formulations remain limited to narrow ribbons and static settings. They do not provide a general dynamic theory applicable to developable surfaces of arbitrary width, which is required for large-scale applications such as space antennas. 

In this work, we remove the restriction on the panel width and assume the panels are infinitely thin. We strictly enforce the developability (isometry) constraint so that the Wunderlich functional can describe the bending energy accurately.  The basic idea is that, the (dynamic) configuration of the developable surface is still governed by the (dynamic) shape of a reference curve. And the shape evolution of the curve is driven by the Wunderlich bending energy.
Following this approach,  we eventually get an ``exact" 1D rod model for the reference curve, by employing the exact 2D Wunderlich energy. The model is a 1D model and can describe the dynamics of a developable surface. Thus, we refer to this formulation as an exact dimension-reduced rod model (or simply “rod model’’ in the main text) for the dynamics of developable surfaces. 

To this end, we begin with the kinematics to establish the correlations between the dynamic motions of the reference curve and the corresponding developable surface. Specifically, we analyze how a moving generatrix influences the surface velocity distribution and derive an explicit kinematic description of the velocity field regardless of the panel width and deformation. We thus derive the kinetic energy.  Combined with the bending energy derived in statics, 
we then have the Lagrangian of the dynamic system. We find that the Lagrangian cannot be explicitly expressed as a functional of curvature $\kappa$ and torsion $\tau$, which prevents us from directly applying variational principles to derive the Euler-Lagrange equations of the system (as equations of $\kappa$ and $\tau$), as commonly done in static theories \citep{dias2012geometric,WEN2024105829}. Instead, leveraging the new Lagrangian, we introduce the panel-width effect, equivalent momentum and angular momentum into the nonlinear static rod model, thereby developing a nonlinear dynamic rod model for developable surfaces with arbitrary reference and deformed configurations. Solving the new dynamical rod equations yields the motions of the reference curve and the developable surfaces under strict isometric constraint. 
{The comparison with classical Kirchhoff rod and plate models is presented to highlight the advantages of the proposed formulation.}
We then extend our analysis to dynamic curve-fold origami. Composed of two developable surfaces joined by a curved fold, the kinematics of curve-fold origami is established similarly. By considering the dynamics of both panels and their geometric and mechanical correlations, we develop a nonlinear dynamical bi-rod model, which enables the study of the dynamics of a dynamic curve-fold origami. To validate the proposed dynamical theories, we conduct theoretical analyses on four representative cases: a rectangular ribbon, a curve-folded plate, a multi-curve-fold origami with vertices, and a folded ring origami. The theoretical results are validated by finite element analysis, demonstrating good agreement between the theory and simulation.

The paper is organized as follows. In Section \ref{sec2}, we start with the kinematics of a single developable surface and then develop an exact dimension-reduced dynamical theory for a single developable surface with arbitrary reference and deformed shapes. The theory relies on the 1D nature of developable surfaces.
We then extend the dynamic theory to a nonlinear dynamical bi-rod model for curve-fold origami by analyzing the panels coupled by the folds in Section \ref{sec3new}. Based on the theory, some illustrative dynamic developable structures are theoretically studied in Section \ref{sec3}\label{mathrefs}, which are validated by finite element analysis. Finally, Section \ref{sec:conclusion} concludes the main points of this paper.

\section{Dynamics of elastic developable surfaces}\label{sec2}
\subsection{Preliminaries} \label{sec2.1}
Before introducing the kinematics, we first recall the geometric variables that determine the shape of a developable surface. In differential geometry, a developable surface is described by 
\begin{equation}\label{e1}
\mathbf{r}(S,v)=\mathbf{r}_0(S)+v \mathbf{l}(S),
\end{equation}
with the developability constraint
\begin{equation} \label{e2}
\mathbf{r}_0^{\prime} \cdot (\mathbf{l} \times \mathbf{l}^{\prime})=0.
\end{equation}
Here $\mathbf{r}_0(S)$ is a selected reference curve on the surface, and $\mathbf{l}(S)$ is the generator at the arclength $S$. In the following analysis, we define the arclength derivative by $\partial_S \mathbf{f}=\mathbf{f}^{\prime}$ and the time derivative by $\partial_t \mathbf{f}=\dot{\mathbf{f}}$ for an arbitrary variable $\mathbf{f}$. Due to developability (isometric constraint), the geometry of the selected reference curve determines the shape of the entire developable surface \citep{do1976differential}, making the system indeed one-dimensional. 

We use the Frenet frame to describe the geometry of the reference curve. Frenet frame is a moving frame only associated with the curve, which consists of the unit tangent $\mathbf{t}$, the unit normal $\mathbf{n}$, and the unit binormal $\mathbf{b}=\mathbf{t} \times \mathbf{n}$ to the curve. The derivatives of the vectors along the curve are
\begin{equation}\label{e3}
\begin{aligned}
     \mathbf{t}^{\prime}&=\kappa \mathbf{n},\\  
     \mathbf{n}^{\prime}&=-\kappa \mathbf{t} + \tau \mathbf{b},\\ 
     \mathbf{b}^{\prime}&=-\tau \mathbf{n},
\end{aligned}
\end{equation}
where $\kappa$ and $\tau$ are the curvature and torsion, which determine the shape of the space curve uniquely up to rigid motions by the fundamental theorem of curves \citep{do1976differential}.

We use the Darboux frame to describe the geometry of the developable surface. Darboux frame is a moving frame associated with both the curve and the surface, which consists of $\mathbf{e}_1$, $\mathbf{e}_2$, $\mathbf{e}_3$ where $\mathbf{e}_1$, $\mathbf{e}_2$ are on the tangent plane and $\mathbf{e}_3$ is the unit normal to the surface. For a developable surface, we use
\begin{equation}\label{e4}
\begin{aligned}
     \mathbf{e}_1&=\mathbf{t},\\  
     \mathbf{e}_3&=\frac{\mathbf{e}_1 \times \mathbf{l}}{\left|\mathbf{e}_1 \times \mathbf{l}\right|},\\ 
     \mathbf{e}_2&=\mathbf{e}_3 \times \mathbf{e}_1,
\end{aligned}
\end{equation}
to connect these two frames, and the $\{\mathbf{e}_1,\mathbf{e}_2,\mathbf{e}_3\}$ frame will not change its direction when moving along a generator. Thus, the Darboux vectors only relate to the arc length $S$, following
\begin{equation}\label{e5}
\begin{aligned}
    \mathbf{e}^{\prime}_i=\bm{\omega} \times \mathbf{e}_i,
\end{aligned}
\end{equation}
where $\bm{\omega}$ is the Darboux vector representing the rotation of the Darboux frames along the curve.
The geodesic curvature $\kappa_g$ and normal curvature $\kappa_n$ of the curve on the surface are defined as
\begin{equation}\label{e7}
\begin{aligned}
    \kappa_g&=\kappa \mathbf{n} \cdot \mathbf{e}_2, \\
    \kappa_n&=\kappa \mathbf{n} \cdot \mathbf{e}_3. 
\end{aligned}
\end{equation}
The geodesic curvature remains constant during isometric deformation, thus it only depends on the reference planar configuration. To describe the distribution of generators, we define
\begin{equation} \label{e8}
\eta=\cot \left\langle\mathbf{e}_1, \mathbf{l}\right\rangle,
\end{equation}
where $\left\langle\mathbf{e}_1, \mathbf{l}\right\rangle$ represents the angle between the vectors $\mathbf{e}_1$ and $\mathbf{l}$, ranging from $0$ to $\pi$. The generator vector $\mathbf{l}$ is then given by
\begin{equation} \label{e9}
\mathbf{l}=\eta \mathbf{e}_1+\mathbf{e}_2.
\end{equation}
According to Eqs. \eqref{e2} - \eqref{e9}, $\bm{\omega}$ is given by
\begin{equation}\label{e10}
\begin{aligned}
    \bm{\omega}=-\eta\kappa_n\mathbf{e}_1-\kappa_n\mathbf{e}_2+\kappa_g\mathbf{e}_3
\end{aligned}
\end{equation}
and $\tau$ can thus be expressed as
\begin{equation}\label{e12}
    \tau=\frac{\mathbf{t} \cdot (\mathbf{t}^{\prime}\times \mathbf{t}^{\prime \prime})}{\kappa^2}=\frac{\mathbf{e}_1 \cdot (\mathbf{e}_1^{\prime}\times \mathbf{e}_1^{\prime \prime})}{\kappa^2}=-\eta \kappa_n+\frac{\kappa_g}{\kappa^2}\kappa_n^{\prime}-\frac{\kappa_n}{\kappa^2}\kappa_g^{\prime},
\end{equation}
indicating that the geometry of the reference curve on the deformed surface, i.e., $\kappa$ and $\tau$, determines the generator distribution and thus determines the shape of the deformed surface.

\begin{figure}[!t]
    \centering
    \includegraphics[width=1\textwidth]{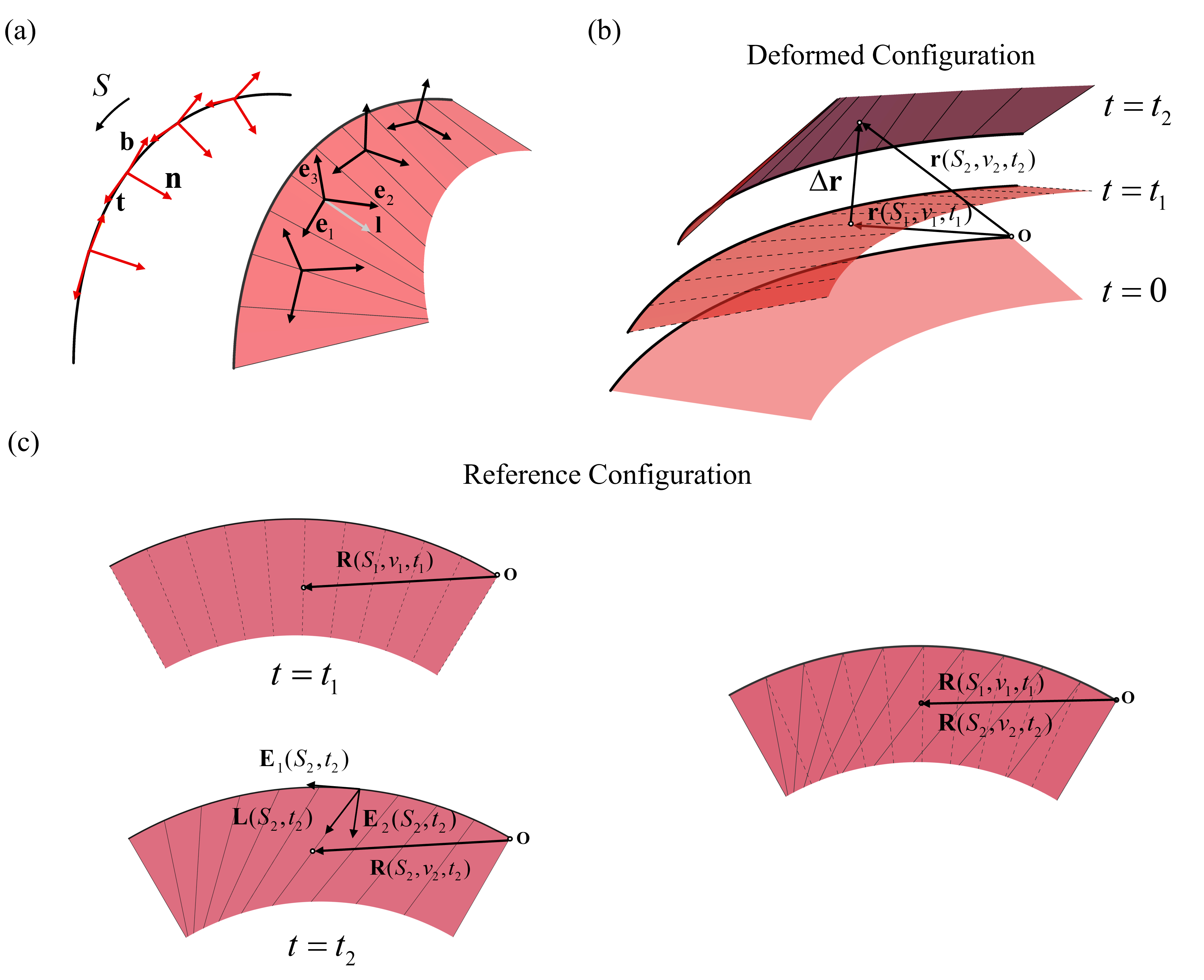}
    \caption{\label{velofig} Kinematics of a developable surface.
(a) Frenet and Darboux frames.
(b) Motion of a material point from $t_1$ to $t_2$ in the deformed configuration.
(c) Motion of a material point from $t_1$ to $t_2$ in the reference configuration.}
\end{figure}

\subsection{Kinematics of developable surfaces}
\label{sec:kinematics}

Having established the preliminary differential geometry framework for developable surfaces, we can now begin the kinematic analysis. Specifically, the kinematic analysis aims to express the velocity distribution of the developable surface in terms of the motion of the reference curve, i.e., in terms of $\mathbf{r}_0$, $\kappa$, and $\tau$. Although Eq. \eqref{e1} gives the position vector of point $(S,v)$ at time $t$, the velocity of the point is not equal to $\partial_t\mathbf{r}$. The reason is that the generator distribution is time-dependent, resulting in changes of the coordinates of the point, as illustrated in Fig. \ref{velofig}(b). Instead, considering the variation of the {generatrix} (frames formed by generators), the velocity of the point is expressed as
\begin{equation}\label{evelo}
    \mathbf{u}(S,v,t)=\frac{\d}{\d t}\mathbf{r}(S,v,t)=\partial_t\mathbf{r}(S,v,t)+\partial_S\mathbf{r}(S,v,t)\frac{\d S}{\d t}+\partial_v\mathbf{r}(S,v,t)\frac{\d v}{\d t}.
\end{equation}
We define $\mathbf{r}(S,v,t)$ as the position vector of point $(S,v)$ at time $t$ in the deformed configuration and $\mathbf{R}(S,v,t)$ the corresponding point in the reference planar configuration. The point preserves its position in the reference configuration, even though it has different $(S,v)$ coordinates in deformed configurations at different time $t$, as shown in Fig. \ref{velofig}(c). Since $\mathbf{R}(S,v,t)$ is preserved, the constraint is expressed as 
\begin{equation} \label{erefc}
    \frac{\d}{\d t}\mathbf{R}(S,v,t)=0,
\end{equation}
where $\mathbf{R}(S,v,t)$ follows the equation for ruled surfaces 
\begin{equation}\label{erefR}
    \mathbf{R}(S,v,t)=\mathbf{R}_0(S,t)+v\mathbf{L}(S,t),
\end{equation}
where $\mathbf{R}_0$ represents the position of the reference curve and $\mathbf{L}$ represents the generator in the reference configuration, which follows $\mathbf{L}=\eta\mathbf{E}_1+\mathbf{E}_2$ and $\mathbf{R}^{\prime}_0=\mathbf{E}_1$ as illustrated in Fig. \ref{velofig}(c). Eq. \eqref{erefc} is thus transformed into
\begin{equation}\label{erefc2}
    \mathbf{R}^{\prime}_0\frac{\d S}{\d t}+\mathbf{L}\frac{\d v}{\d t}+v(\mathbf{L}^{\prime}\frac{\d S}{\d t}+\dot{\mathbf{L}})=0.
\end{equation}
Solving Eq. \eqref{erefc2} yields:
\begin{equation}\label{eqsv}
\begin{aligned}
    &\frac{\d S}{\d t}=\frac{v \dot{\eta}}{1-v(1+\eta^2)\kappa_g+v\eta^{\prime}},\\
    &\frac{\d v}{\d t}=-v\eta\kappa_g\frac{\d S}{\d t}.
\end{aligned}
\end{equation}
Substituting Eq. \eqref{eqsv} into Eq. \eqref{evelo}, we have the velocity
\begin{equation}\label{equ1}
    \mathbf{u}=\dot{\mathbf{r}}_0+v(\eta\dot{\mathbf{e}}_1+\dot{\mathbf{e}}_2).
\end{equation}
The angular velocity of the Darboux vector $\bm\Omega$ is defined as
\begin{equation}
    \dot{\mathbf{e}}_i=\bm\Omega \times \mathbf{e}_i.
\end{equation}
Eq. \eqref{equ1} can thus be written in a rather concise way
\begin{equation}\label{equ2}
    \mathbf{u}=\dot{\mathbf{r}}_0+ v\bm\Omega \times \mathbf{l},
\end{equation}
which indicates that the instantaneous motion of point $(S,v)$ consists of the superposition of the translational velocity of point $(S,0)$ and the rotational velocity component due to the angular velocity $\bm\Omega$ about point $(S,0)$. According to Eq. \eqref{equ2} and Eq. \eqref{eqsv}, we can further obtain the distribution of the acceleration
\begin{equation}\label{eacc}
\begin{aligned}
    \frac{\d\mathbf{u}}{\d t}&=\ddot{\mathbf{r}}_0+\dot{\mathbf{r}}^{\prime}_0\frac{\d S}{\d t}+v\dot{\bm{\Omega}}\times\mathbf{l}+v\bm{\Omega}\times(\dot{\mathbf{l}}+\mathbf{l}^{\prime}\frac{\d S}{\d t})+\frac{\d v}{\d t}\bm{\Omega}\times\mathbf{l}+v\frac{\d S}{\d t}\bm{\Omega}^{\prime}\times\mathbf{l}\\
    &=\ddot{\mathbf{r}}_0+v\dot{\bm{\Omega}}\times\mathbf{l}+v\bm{\Omega}\times(\bm{\Omega}\times\mathbf{l})+v\frac{\d S}{\d t}\bm{\Omega}^{\prime}\times\mathbf{l}.
\end{aligned}
\end{equation}
The first three terms in Eq. \eqref{eacc} represent the translational acceleration, rotational acceleration, and centripetal acceleration, similar to the acceleration in a rigid body, while the fourth term is an acceleration induced by frame variations.

\subsection{Lagrangian of developable surfaces: kinetic and elastic energy}\label{sec2.2}

The Lagrangian of a dynamical system, consisting of the elastic and kinetic energy, determines its motion. For an elastic developable surface, the elastic energy density only contains the bending energy term, which is proportional to the square of its nonzero principal curvature $\kappa_m$. Given by the \citet{Wunderlich1962} functional, the elastic energy $V$ of a developable surface is
 \begin{equation} \label{ewunder}
 \begin{aligned}
 V=\frac{1}{2}D \int_0^{S_0}
  \frac{\left(1+\eta^2\right)^2 \kappa_{n}^2}{\eta^{\prime}-\left(1+\eta^2\right) \kappa_{g}} \ln [1+ \eta^{\prime}v_{0}-(1+\eta^2) \kappa_{g}v_{0}] \d S=\int_0^{S_0}\varepsilon \d S,
 \end{aligned} 
\end{equation}
where $D$ is the bending stiffness and $\varepsilon$ is the line density of elastic energy along the reference curve. Eq. \eqref{ewunder} indicates that $V$ is determined by the curvature of the reference curve and the generator distribution (or determined by the curvature and torsion of the reference curve).

Based on the kinematic analysis in Section \ref{sec:kinematics}, we derive the kinetic energy $T$ as
\begin{equation} \label{eT1}
    \begin{aligned}
        T=\frac{1}{2}\rho h \int_0^{S_0}\int_0^{v_0} \left|\mathbf{J}\right|\mathbf{u}\cdot\mathbf{u} \d v \d S,
    \end{aligned}
\end{equation}
where $\rho$ is the mass density, $h$ is the surface thickness and $\mathbf{J}$ is the Jacobian matrix following
\begin{equation} \label{ejaco}
    \left|\mathbf{J}\right|=\left|\partial_S \mathbf{r} \times \partial_v \mathbf{r}\right|=1+ \eta^{\prime}v-(1+\eta^2)\kappa_{g}v.
\end{equation}
Substituting Eq. \eqref{equ1} and \eqref{ejaco} into Eq. \eqref{eT1} yields
\begin{equation}
    T=\int_0^{S_0}\lambda \d S,
\end{equation}
where $\lambda$ is the kinetic energy line density with
\begin{equation}\label{evareT}
    \begin{aligned}
        \lambda&=\frac{1}{2}\rho h v_0((1+\frac{1}{2}v_0(\eta'-(1+\eta^2)\kappa_g)\dot{\mathbf{r}}_0\cdot\dot{\mathbf{r}}_0+(\frac{1}{2}v_0+\frac{1}{3}v_0^2(\eta'-(1+\eta^2)\kappa_g))\dot{\mathbf{r}}_0\cdot(\bm{\Omega}\times\mathbf{l})\\&+(\frac{1}{3}v_0^2+\frac{1}{4}v_0^3(\eta'-(1+\eta^2)\kappa_g))(\bm{\Omega}\times\mathbf{l})\cdot(\bm{\Omega}\times\mathbf{l})).
    \end{aligned}
\end{equation}
Eqs. \eqref{ewunder} and \eqref{evareT} indicate that the Lagrangian, i.e., $L=T-V$, can be expressed as a one-dimensional integral along the reference curve. In statics analysis, we have $L=-V$, which is a functional of $\kappa$ and $\tau$, and the Lagrangian can directly lead to the Euler-Lagrange equations in terms of $\kappa$ and $\tau$ \citep{dias2012geometric,WEN2024105829}. However, in dynamic analysis, the variables including $\mathbf{r}_0$ and $\bm\Omega$ cannot be explicitly expressed in terms of $\kappa$ and $\tau$, resulting in a complex Lagrangian that cannot directly lead to the governing equations. To address this issue, we draw inspiration from the equivalent rod model for static developable surfaces \citep{DIAS201457,dias2015wunderlich} with narrow ribbons,
and develop an equivalent model to handle the dynamical analysis for a general developable surface.

\subsection{A 1D nonlinear dynamical model for general developable surfaces} \label{sec2.3}
As a conclusion of Section \ref{sec2.2}, the Lagrangian of a dynamic developable surface is a one-dimensional integral along the crease. Therefore, we can always establish a non-linear 1D dynamical model (similar to a dynamic rod model) with the same Lagrangian. The key point is that the two systems must have the same Lagrangian, which ensures that the motion of the reference curve on the developable surface matches that of the effective rod predicted by the 1D dynamic model. In this section, we derive explicit dynamical equations for the reference curve, similar to a nonlinear rod model. The new rod model will give the dynamics of the reference curve and thus the dynamics of the developable surface under isometric constraint. Since we use the Wunderlich functional (no narrow ribbon assumption), our model is an exact dimension-reduced model that can predict the dynamics of a general developable surface accurately. 

\begin{figure}[t]
    \centering
    \includegraphics[width=1\textwidth]{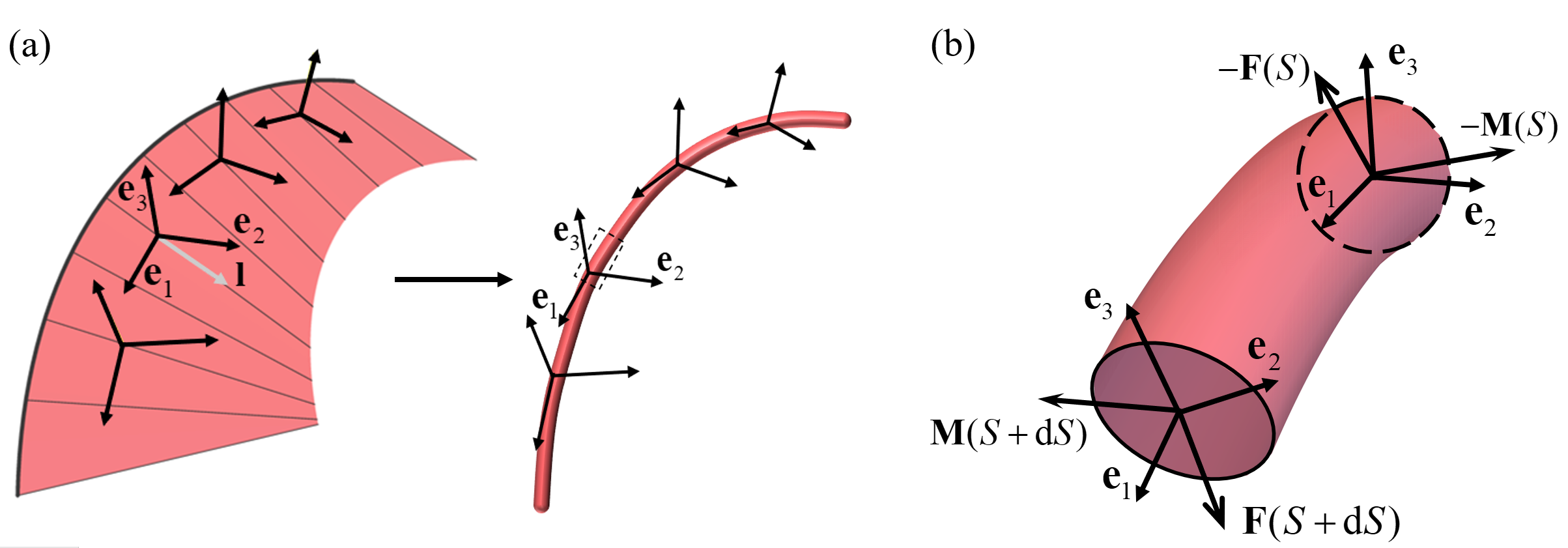}
    \caption{\label{fig:sec2.3} (a) Correspondence between the local frames of a developable surface and those of the equivalent rod. (b) Force diagram for a differential rod element.}
\end{figure}
\subsubsection{Constitutive laws of the effective rod}
As shown in Fig.~\ref{fig:sec2.3}, we model the reference curve on a developable surface as an effective rod, using the elastic energy given by Eq.~(\ref{ewunder}).
In the effective rod model, the nonlinear bending and twisting energies determine the total elastic energy. By selecting the $\{\mathbf{e}_2,\mathbf{e}_3\}$  plane as the equivalent rod's cross-section, $\omega_1$ and $\omega_2$ represent the twisting and the bending, respectively, in two directions, while $\omega_3$ is constrained by the geodesic curvature. The variation of the elastic energy in Eq. \eqref{ewunder} is
\begin{equation}\label{eqvari1}
\begin{aligned}
        \delta V&=\int \delta \varepsilon \d S=\int\left(\frac{\partial \varepsilon}{\partial \kappa_n} \delta \kappa_n+\left(-\frac{\d}{\d S} \frac{\partial \varepsilon}{\partial \eta^{\prime}}+\frac{\partial \varepsilon}{\partial \eta}\right) \delta \eta\right) d S+\left.\frac{\partial \varepsilon}{\partial \eta^{\prime}} \delta \eta\right|_0 ^{S_0}\\
        &=\int(\frac{1}{\kappa_n}\left(\frac{\d}{\d S} \frac{\partial \varepsilon}{\partial \eta^{\prime}}-\frac{\partial \varepsilon}{\partial \eta}\right) \delta\omega_1+\left(-\frac{\partial \varepsilon}{\partial \kappa_n}-\frac{\eta}{\kappa_n}(\frac{\d}{\d S} \frac{\partial \varepsilon}{\partial \eta^{\prime}}-\frac{\partial \varepsilon}{\partial \eta})\right) \delta\omega_2)\d S+ \left.\frac{\partial \varepsilon}{\partial \eta^{\prime}} \delta \eta\right|_0 ^{S_0}
\end{aligned}
\end{equation}
where the integral is defined from the starting point to the ending point of the crease, and this definition will be applied in subsequent discussions. Meanwhile, the variation of the internal work in the rod is 
\begin{equation}
    \delta V^*=\int( m_1\delta\omega_1+m_2\delta\omega_2+m_3\delta\omega_3 )\d S,
\end{equation}
where $m_i$ is the component form of the internal moment $\mathbf{M}$, i.e., the moment of the part $(S,S_0)$ acting on the other part $(0,S)$. Notice that the geodesic curvature remains constant in isometric deformation, resulting in $\delta\omega_3=0$ and $m_3$ is thus a constraint moment irrelevant to the rods' geometry.  To ensure that the variations of the elastic energy for both systems are equal, the internal moment of the equivalent rod follows the constitutive law {
\begin{equation} \label{em1}
    \mathbf{M}=\frac{1}{\kappa_n}\left(\frac{\d}{\d S} \frac{\partial \varepsilon}{\partial \eta^{\prime}}-\frac{\partial \varepsilon}{\partial \eta}\right) \mathbf{e}_1+\left(-\frac{\partial \varepsilon}{\partial \kappa_n}-\frac{\eta}{\kappa_n}(\frac{\d}{\d S} \frac{\partial \varepsilon}{\partial \eta^{\prime}}-\frac{\partial \varepsilon}{\partial \eta})\right) \mathbf{e}_2+m_3 \mathbf{e}_3.
\end{equation}}
Besides, the equilibrium requires 
\begin{equation}\label{eqbd}
\left.\frac{\partial \varepsilon}{\partial \eta^{\prime}} \delta \eta\right|_0 ^{S_0}=0
\end{equation}
to eliminate the boundary values in Eq. \eqref{eqvari1}. In the narrow ribbon limit, i.e., $v_0\kappa_g\rightarrow0$, the Wunderlich functional degenerates to the \citet{sadowsky1930theorie} functional, and Eq. \eqref{em1} is simplified into 
\begin{equation}
    \begin{aligned}
        \mathbf{M}&=-2Dv_0 \eta\kappa_n(1+\eta^4)\mathbf{e}_1-Dv_0 \kappa_n(1-\eta^2)\mathbf{e}_2+m_3\mathbf{e}_3,\\
    \end{aligned}
\end{equation}
which is the same as Eq. (12) in \citep{DIAS201457}, indicating that our general rod model is consistent with the simplified one. 

\subsubsection{Dynamical equations}
With the constitutive law, the dynamical equations for the effective rod are
\begin{equation} \label{eqdynamic}
    \begin{aligned}
        &\mathbf{F}^{\prime}=\frac{\d \mathbf{p}}{\d t},\\
        &\mathbf{M}^{\prime}+\mathbf{r}^{\prime}_0\times \mathbf{F}=\frac{\d \mathbf{p}_{\theta}}{\d t},\\
        &\mathbf{M}=\frac{1}{\kappa_n}\left(\frac{\d}{\d S} \frac{\partial \varepsilon}{\partial \eta^{\prime}}-\frac{\partial \varepsilon}{\partial \eta}\right) \mathbf{e}_1+\left(-\frac{\partial \varepsilon}{\partial \kappa_n}-\frac{\eta}{\kappa_n}(\frac{\d}{\d S} \frac{\partial \varepsilon}{\partial \eta^{\prime}}-\frac{\partial \varepsilon}{\partial \eta})\right) \mathbf{e}_2+m_3 \mathbf{e}_3,
    \end{aligned}
\end{equation}
where $\mathbf{F}$ is the internal force. This section aims to derive the equivalent momentum $\mathbf{p}$ and angular momentum $\mathbf{p_{\theta}}$ (with respect to the fixed reference point with the same coordinate as $\mathbf{r}_{0}(S,t)$) of the effective rod per unit length. The derivation is carried out in the Lagrangian framework, where the material segment between S and S + $\mathrm{d}S$ is fixed and only velocity derivatives appear in Eq.~\eqref{eqmom}. In contrast, an Eulerian formulation must account for both mass and velocity transport across $S$ and $S + \mathrm{d}S$ arising from the evolution of the generator distribution. As shown in \ref{AppA}, these two viewpoints lead to equivalent expressions for the momentum and angular momentum.

As stated in Section \ref{sec:kinematics}, the instantaneous motion of the point $(S,v)$ consists of the superposition of the translational velocity of the point $(S,0)$ and the rotational velocity component due to the angular velocity $\bm\Omega$ about the point $(S,0)$. Thus, the derivative of the momentum and angular momentum line density of the rod element is
\begin{equation}\label{eqmom}
\begin{aligned}
       &\frac{\d \mathbf{p}}{\d t}=\rho h \int^{v_0}_0 \left|\mathbf{J}\right|\frac{\d\mathbf{u}}{\d t}\d v,\\
       &\frac{\d \mathbf{p}_\theta}{\d t}=\rho h \int^{v_0}_0\left|\mathbf{J}\right|v\mathbf{l}\times\frac{\d\mathbf{u}}{\d t}\d v.
\end{aligned}
\end{equation}
Substituting Eqs. \eqref{eqmom} and \eqref{eacc} into \eqref{eqdynamic}, the full dynamical equation system is 
\begin{equation} \label{eqdynamicfull}
    \begin{aligned}
         &\mathbf{F}^{\prime}=\rho h [(v_0+\frac{1}{2}v_0^2\gamma)\ddot{\mathbf{r}}_0+(\frac{1}{2}v_0^2+\frac{1}{3}v_0^3\gamma)(\dot{\bm{\Omega}}\times\mathbf{l}+\bm{\Omega}\times(\bm{\Omega}\times\mathbf{l}))-\frac{1}{3}v_0^3\dot{\eta}\bm{\Omega}^{\prime}\times\mathbf{l}],\\
        &\mathbf{M}^{\prime}+\mathbf{r}^{\prime}_0\times \mathbf{F}=\rho h[(\frac{1}{2}v_0^2+\frac{1}{3}v_0^3\gamma)\mathbf{l}\times\ddot{\mathbf{r}}_0+(\frac{1}{3}v_0^3+\frac{1}{4}v_0^4\gamma)\mathbf{l}\times(\dot{\bm{\Omega}}\times\mathbf{l}+\bm{\Omega}\times(\bm{\Omega}\times\mathbf{l}))-\frac{1}{4}v_0^4\dot{\eta}\mathbf{l}\times(\bm{\Omega}^{\prime}\times\mathbf{l})],\\
        &\mathbf{M}=\frac{1}{\kappa_n}\left(\frac{\d}{\d S} \frac{\partial \varepsilon}{\partial \eta^{\prime}}-\frac{\partial \varepsilon}{\partial \eta}\right) \mathbf{e}_1+\left(-\frac{\partial \varepsilon}{\partial \kappa_n}-\frac{\eta}{\kappa_n}(\frac{\d}{\d S} \frac{\partial \varepsilon}{\partial \eta^{\prime}}-\frac{\partial \varepsilon}{\partial \eta})\right) \mathbf{e}_2+m_3 \mathbf{e}_3,
    \end{aligned}
\end{equation}
where $\gamma=\eta^{\prime}-(1+\eta^2)\kappa_g$ and $m_3$ is a Lagrangian multiplier constrained by the geodesic curvature. The dynamic equation system (\ref{eqdynamicfull}) is one of the key contributions in this work. It employs the Wunderlich functional for wide developable surfaces and extends the well-studied statics to the dynamic regime. The left-hand side of (\ref{eqdynamicfull}) considers the force/moment equilibrium along the effective rod, and the right-hand side contains the time evolution terms, resulting in the shape change with respect to time. The numerical scheme to solve the spatial-temporal evolution of (\ref{eqdynamicfull}) is given in \ref{AppB}.

As stated previously, Eqs.~\eqref{eqdynamicfull} are derived within the Lagrangian framework. In \ref{AppA}, we prove that the result is equivalent in the Eulerian framework. At both ends of the rod, the deformation satisfies the boundary conditions
\begin{equation}\label{eqbvp_de}
    \begin{aligned}
        &\mathbf{F}=\mathbf{F}_e,\\
        &\mathbf{M}=\mathbf{M}_e.
    \end{aligned}
\end{equation}
Eqs. \eqref{eqdynamicfull} and \eqref{eqbvp_de} characterize the full dynamical evolution of the reference curve, whose motion in turn determines the deformation of the entire developable surface. 

\subsection{Model validation: dynamics of a lifted rolled ribbon}\label{seckrod}

\begin{figure}[!b]
    \centering
    \includegraphics[width=1\textwidth]{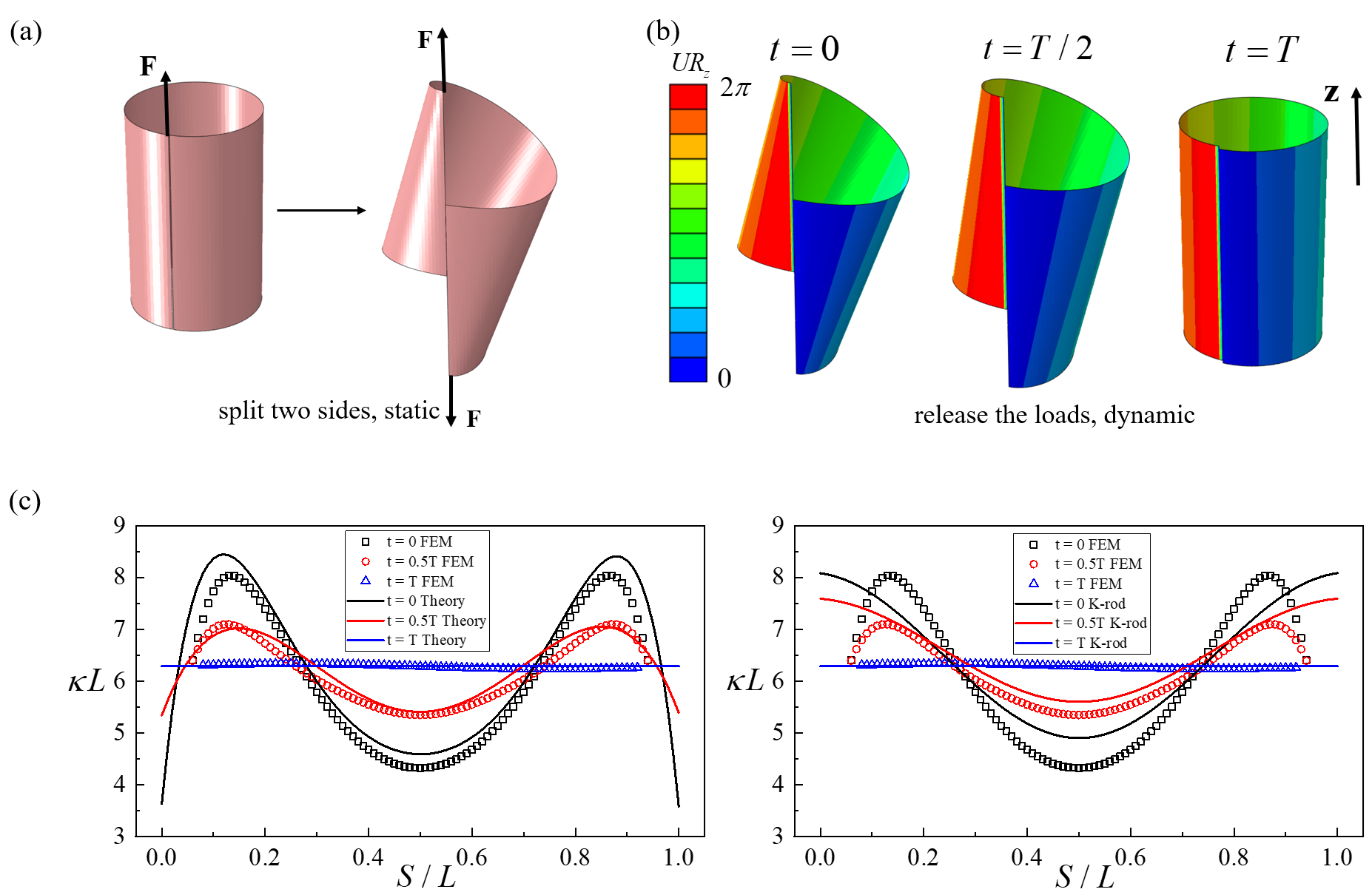}
    \caption{\label{compare} \comment{(a) A rectangular wide strip is rolled and lifted to the initial position. {(b) The dynamic process after releasing the load. The color bar denotes the rotation angle around $z$ axis.} (c) The quantitative results (nondimensionalized curvature $\kappa L$) predicted by our model (left) and the Kirchhoff rod model (right). Our model achieves better agreement with the FEM results. $T$ is the characteristic evolution time shown in (b).}}
\end{figure}

In this section, we compare the results of our new model and the Kirchhoff rod model in the example shown in Fig. \ref{compare}. As illustrated, a rectangular ribbon with aspect ratio $L/b=2$ is first rolled and split statically, and we study the dynamic process after the lifting load is released.

For a rectangular ribbon, if both wide sides remain straight, the projection of the generator length $v_0$ remains constant, which equals the rectangular width. Under such circumstances, it is proper to select the centerline as the reference curve due to symmetry, instead of the boundary line. For the centerline, Eq.~\eqref{eqdynamicfull} is transformed into

\begin{equation} \label{eqdynamicfull_centerline}
    \begin{aligned}
         &\mathbf{F}^{\prime}=\rho h [v_0\ddot{\mathbf{r}}_0+\frac{1}{12}v_0^3\gamma(\dot{\bm{\Omega}}\times\mathbf{l}+\bm{\Omega}\times(\bm{\Omega}\times\mathbf{l}))-\frac{1}{12}v_0^3\dot{\eta}\bm{\Omega}^{\prime}\times\mathbf{l}],\\
        &\mathbf{M}^{\prime}+\mathbf{r}^{\prime}_0\times \mathbf{F}=\rho h[\frac{1}{12}v_0^3\gamma\mathbf{l}\times\ddot{\mathbf{r}}_0+\frac{1}{12}v_0^3\mathbf{l}\times(\dot{\bm{\Omega}}\times\mathbf{l}+\bm{\Omega}\times(\bm{\Omega}\times\mathbf{l}))],\\
        &\mathbf{M}=\frac{1}{\kappa_n}\left(\frac{\d}{\d S} \frac{\partial \varepsilon}{\partial \eta^{\prime}}-\frac{\partial \varepsilon}{\partial \eta}\right) \mathbf{e}_1+\left(-\frac{\partial \varepsilon}{\partial \kappa_n}-\frac{\eta}{\kappa_n}(\frac{\d}{\d S} \frac{\partial \varepsilon}{\partial \eta^{\prime}}-\frac{\partial \varepsilon}{\partial \eta})\right) \mathbf{e}_2+m_3 \mathbf{e}_3,
    \end{aligned}
\end{equation}
where
\begin{equation}
    \varepsilon=\frac{1}{2}D 
  \frac{\left(1+\eta^2\right)^2 \kappa^2}{\eta^{\prime}} \ln (\frac{2+ \eta^{\prime}v_{0}}{2- \eta^{\prime}v_{0}}).
\end{equation}

We also notice that under extreme approximations, i.e., $v_0\to0$, $\tau/\kappa_n\to 0$ and $\kappa_g=0$, Eq.~\eqref{eqdynamicfull} is transformed into
\begin{equation}\label{ekirch}
    \begin{aligned}
        &\mathbf{F}^{\prime}=\rho h v_0\ddot{\mathbf{r}}_0,\\
        &\mathbf{M}^{\prime}+\mathbf{r}^{\prime}_0\times \mathbf{F} =\frac{1}{12}\rho hv_0^3\mathbf{e}_2\times(\dot{\bm{\Omega}}\times\mathbf{e}_2+\bm{\Omega}\times(\bm{\Omega}\times\mathbf{e}_2)),\\
        &\mathbf{M}=2Dv_0 \tau \mathbf{e}_1-Dv_0 \kappa_n\mathbf{e}_2+m_3\mathbf{e}_3,
    \end{aligned}
\end{equation}
along with the boundary conditions $F=0$, $m=0$. Eq. \eqref{ekirch} is equivalent to the dynamical equations of the Kirchhoff rod model \citep{coleman1993dynamics,Goriely2001, HUANG2024105721}, which implies that the model established in this work can degenerate into existing static and dynamical models. In the Kirchhoff rod model (\ref{ekirch}), the curvature $\kappa = \kappa_n$ and torsion $\tau$ of the rod are decoupled, in contrast with the coupling of $\kappa$ and $\tau$ in our model. Fig. \ref{compare} compares these two models to validate our new model's ability to handle highly twisted panels with arbitrary width. The comparison between the theoretical results derived via different models along with the FEM results is shown in Fig.~\ref{compare}(c), illustrating that the Kirchhoff rod model has difficulties handling highly twisted wide panels, while the coupling of curvature and torsion along with the motion of local frames in our theory leads to the accurate modeling of arbitrarily deformed developable surfaces using our model.

{
\subsection{Model validation: transient response of an elastic plate}\label{secKL}
In the field of thin sheets and plates, the Kirchhoff–Love plate model is also a classic and commonly used model. In this section, we compare our new model and the Kirchhoff-Love plate model in the example shown in Fig. \ref{KLcompare}. A plate (length: $L$, width: $b$, thickness: $h$) with simply supported sliding ends is subjected to an instantaneous velocity field  following
\begin{equation}
    \dot{w}(x,y)=0.3\frac{\pi^2}{L^2}\sqrt{\frac{D}{\rho h}}\sin\frac{\pi x}{L}
\end{equation}
where $w$ is the vertical displacement.
The plate
 consequently begins to vibrate, as illustrated in Fig. \ref{KLcompare}(a). Following the $y$ invariance, the dynamic equations Eq. \eqref{eqdynamicfull} are thus simplified into
\begin{equation} \label{platefulleq}
    \begin{aligned}
        &F_w\cos\varphi-F_x\sin\varphi+D\varphi''=0,\\
        &F_x=\rho h\ddot{x},\quad F_w=\rho h\ddot{w},\\
        &x'=\cos\varphi, \quad  w'=\sin\varphi,\\
    \end{aligned}
\end{equation}
along with the boundary conditions 
\begin{equation} \label{platebound}
    F_x=0, \quad\varphi''=0,\quad w=0,
\end{equation}
at both ends. Solving Eqs. \eqref{platefulleq} and \eqref{platebound} yields the motion of the plate. 
\begin{figure}[!b]
    \centering
    \includegraphics[width=1\textwidth]{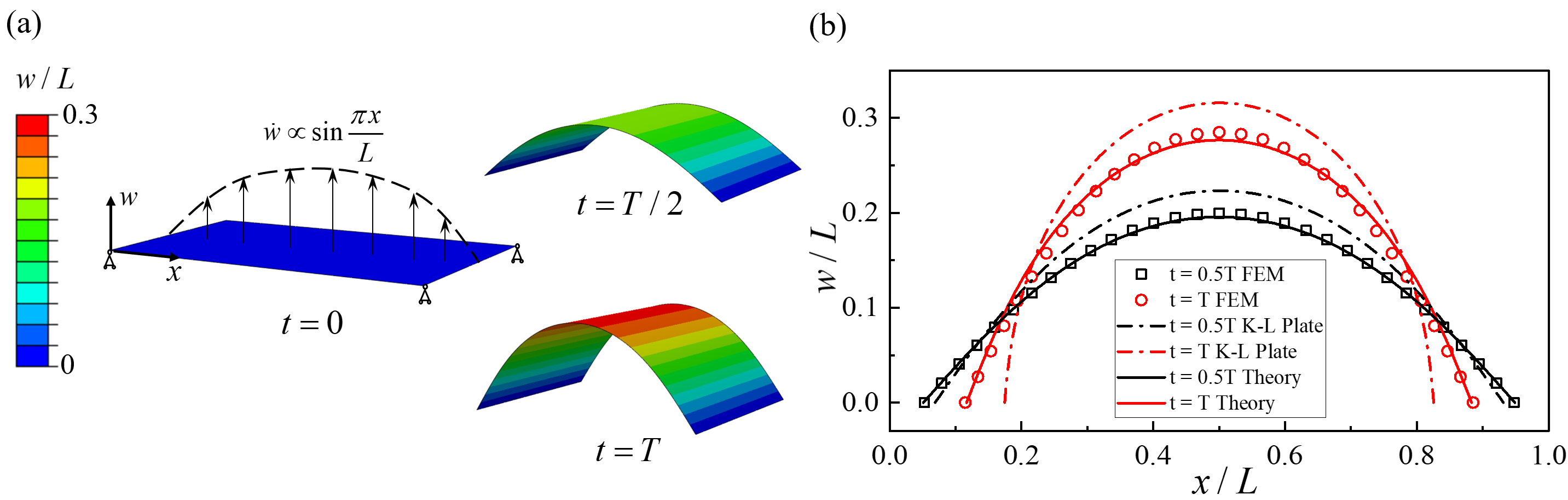}
    \caption{\label{KLcompare} (a) The vibration of an elastic plate with an initial velocity field. (b) The configuration (in the $xw$ plane) predicted by our model and the Kirchhoff-Love plate model. Our model achieves better agreement with the FEM results. The length $L$ is used to nondimensionalize the geometry.}
\end{figure}
In Kirchhoff-Love plate theory, the dynamic equation follows  \citep{Kirchhoff+1850+51+88}
\begin{equation} \label{KLdynamic}
    Dw''''+\rho h\ddot{w}=0.
\end{equation}
With the boundary conditions Eq. \eqref{platebound}, the explicit solution of Eq. \eqref{KLdynamic} is
\begin{equation}
    \begin{aligned}
        w=A\sin\frac{\pi S}{L}\sin (\frac{\pi^2}{L^2}\sqrt{\frac{D}{\rho h}}t),
    \end{aligned}
\end{equation}
where A is the amplitude determined by the initial velocity field.

To validate the two theoretical models, we derive the configuration at time $T$ (defined as the time when the plate first reaches zero velocity) and $0.5T$ from both Kirchhoff-Love plate theory and the new theory. As illustrated in Fig. \ref{KLcompare}(b), the solution of the new theory aligns better with finite element results, indicating that the new theory can handle large curvature problems more accurately. 

As a conclusion of Sections \ref{seckrod} and \ref{secKL}, we summarize the assumptions, advantages and limitations of the Kichhoff-rod model, Kirchhoff-Love plate model and our new model in the following chart. 

\begin{figure}[!h]
    \centering
    \includegraphics[width=0.8\linewidth]{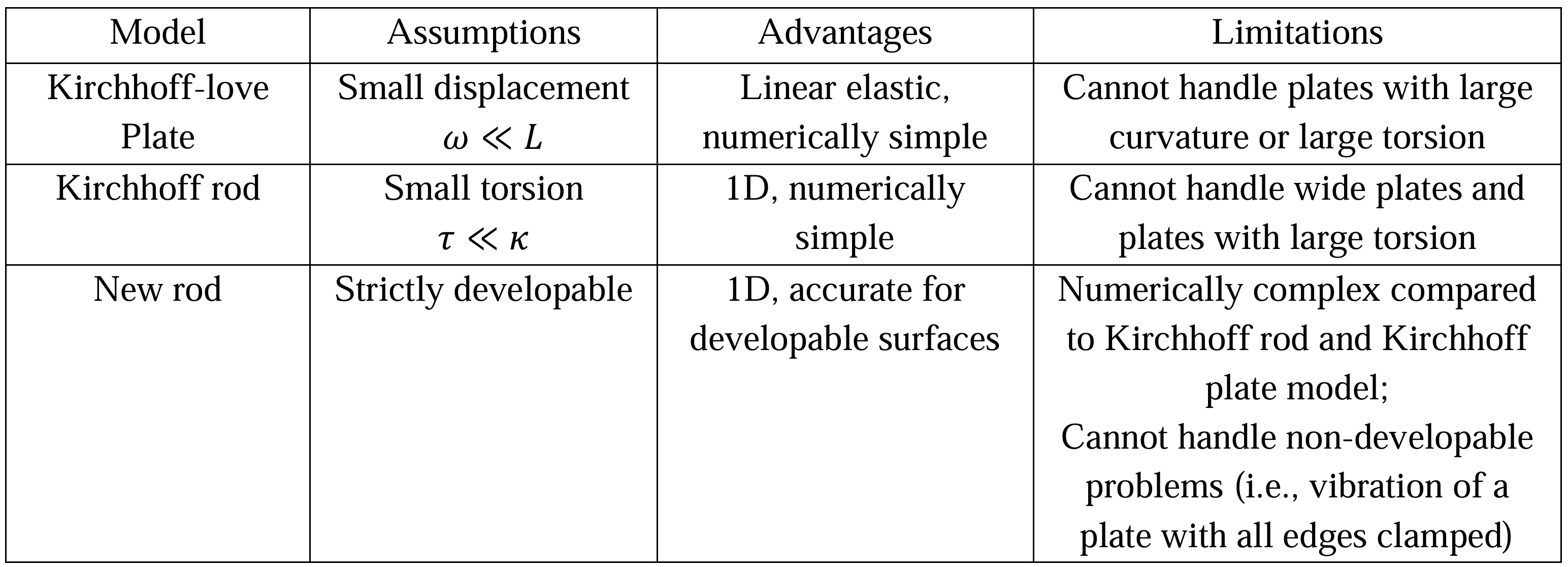}
\end{figure}

In this section, we have therefore established an accurate dynamic theory for developable surfaces—valid for panels of arbitrary width—and recast it into an equivalent rod formulation suitable for numerical implementation. In the next section, we extend this framework from single developable surfaces to curved-fold origami, which consists of two coupled developable panels.
}

\section{Dynamics of curve-fold origami}\label{sec3new}

\subsection{Lagrangian of curve-fold origami}\label{secnew22}
Following similar procedures as in Section \ref{sec2}, we start with the elastic energy of the curve-fold origami structures. 
Curve-fold origami consists of multiple developable surfaces meeting at the creases. The elastic energy of a single-curve-fold origami structure thus follows
\begin{equation}
    V=V_{bend+}+V_{bend-}+V_{fold},
\end{equation}
where $V_{bend\pm}$ represent the bending energies of the two developable surfaces and $V_{fold}$ represents the folding energy of the crease. $V_{fold}$  employs the torsional spring model 
\begin{equation}\label{ettfold}
    V_{fold}=\int \frac{1}{2}k_c(\phi-\phi_0)^2\d S,
\end{equation}
where $k_c$ is the torsional stiffness of a unit-length fold, $\phi$ is the folding angle satisfying
\begin{equation}\label{ephi}
    \phi=2 \arctan(\frac{\kappa_n}{\kappa_g}),
\end{equation}
and $\phi_0$ is the rest angle. 

As stated in Section \ref{sec2.2}, the bending energy of a developable surface is determined by the curvature and torsion of its reference curve. Selecting the crease as the reference curve, the bending energy of the system is 
\begin{equation}\label{ettbend}
\begin{aligned}
   &V_{bend+}+V_{bend-}=\Sigma_{\pm}\frac{1}{2}D \int
  \frac{\left(1+\eta_{\pm}^2\right)^2 \kappa_{n\pm}^2}{\pm\eta_{\pm}^{\prime}\mp\left(1+\eta_{\pm}^2\right) \kappa_{g}} \ln [1\pm \eta_{\pm}^{\prime}v_{0}\mp(1+\eta_{\pm}^2) \kappa_{g}v_{0\pm}] \d S, \\
  &\eta_{\pm}=\frac{1}{\kappa_{n\pm}}(-\tau+\frac{\kappa_g}{\kappa^2}\kappa_{n\pm}^{\prime}-\frac{\kappa_{n\pm}}{\kappa^2}\kappa_{g}^{\prime}).
\end{aligned}
\end{equation}
From Eqs. \eqref{ettfold}, \eqref{ephi} and \eqref{ettbend}, the elastic energy is determined by the curvature $\kappa$ and  torsion $\tau$ of the fold, given by
\begin{equation}
V=\int\varepsilon(\kappa,\kappa^{\prime},\kappa^{\prime\prime},\tau,\tau^{\prime})\d S.
\end{equation}

The kinetic energy, according to Eq. \eqref{evareT}, satisfies
\begin{equation}
    T_{+}+T_{-}=\Sigma_{\pm}\int\lambda_{\pm}\d S=\int\lambda(\dot{\mathbf{r}}_0,\kappa,\kappa^{\prime},\kappa^{\prime\prime},\dot{\kappa},\ddot{\kappa},\dot{\kappa}^{\prime},\tau,\tau^{\prime},\dot{\tau}),
\end{equation}
and the Lagrangian is thus derived by
\begin{equation}
    L=(T_++T_-)-(V_{bend+}+V_{bend-})=\int\varepsilon(\dot{\mathbf{r}}_0,\kappa,\kappa^{\prime},\kappa^{\prime\prime},\dot{\kappa},\ddot{\kappa},\dot{\kappa}^{\prime},\tau,\tau^{\prime},\dot{\tau}) \d S.
\end{equation}

As emphasized in Section~\ref{sec2.2}, the full Lagrangian—which depends explicitly on $\dot{\mathbf{r}}_0$—cannot be used directly to obtain governing equations for the curvature and torsion of the fold. Because a single curved-fold origami consists of two coupled developable surfaces, we extend the nonlinear single-rod formulation of Section~\ref{sec2.3} to a nonlinear bi-rod model to address the dynamic problem in the following section.

\subsection{A nonlinear dynamic bi-rod model for curve-fold origami}\label{sec2.5}

In this section, we theoretically illustrate how curve-fold origami structures can be transformed into a dynamic bi-rod model based on the single-rod model previously developed. 
Since curve-fold origami can be regarded as two developable surfaces joined together at the fold, it is natural to take into account the interaction between the two surfaces and then analyze their dynamics separately. Then, the curve-fold origami is equivalent to a bi-rod by adding the interactions between the rods. 

\begin{figure}[t]
    \centering
    \includegraphics[width=0.7\textwidth]{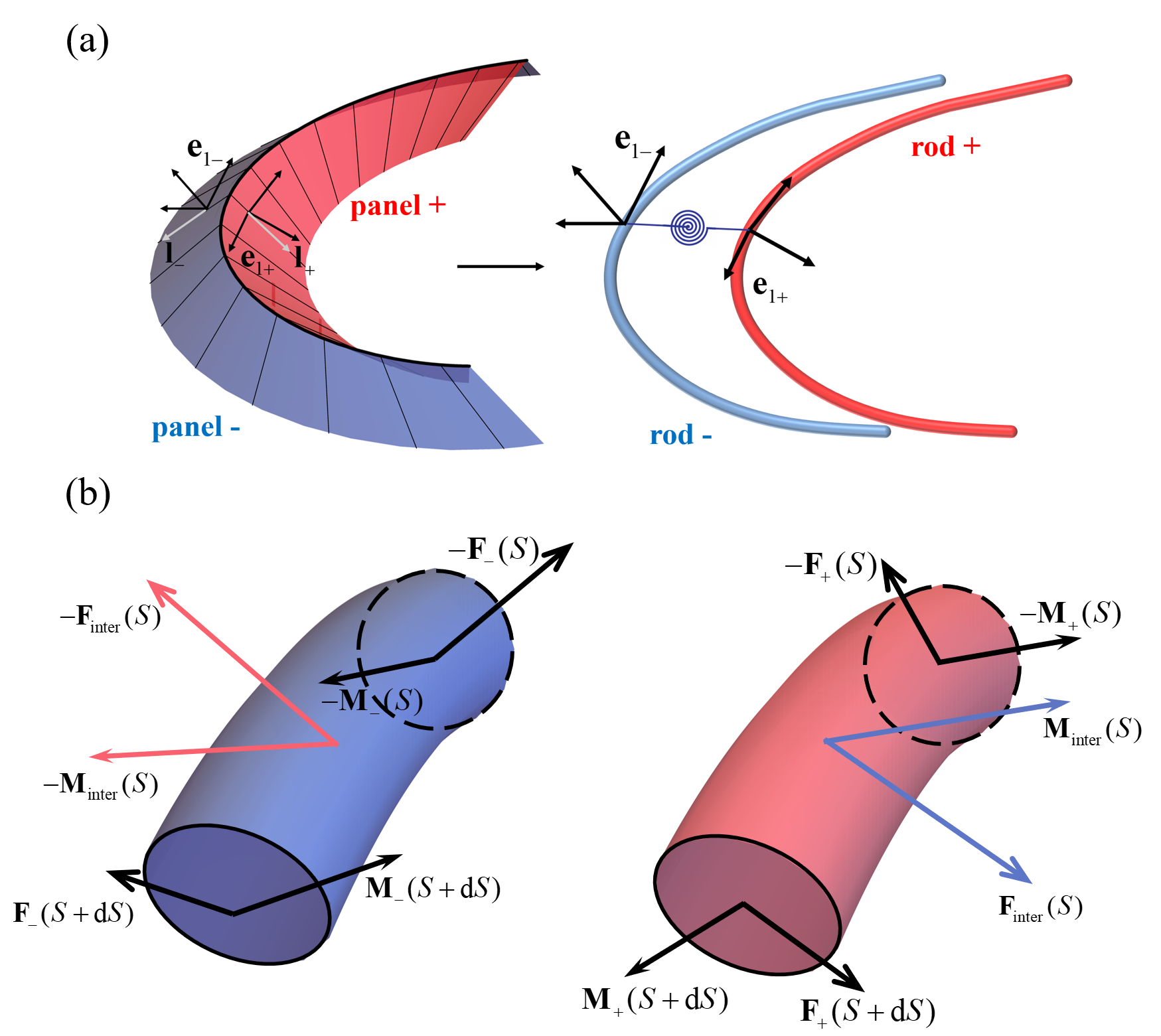}
    \caption{\label{fig:sec24} (a) Correspondence between the local frames of the curved-fold origami and those of the equivalent bi-rod. (b) Force analysis of the corresponding differential rod elements. }
\end{figure}

To derive the  dynamical equations, we first define the Darboux frames of both surfaces on a curve-fold origami associated with the Frenet frame by
\begin{equation}\label{framedtof}
\begin{aligned}
        &\mathbf{e}_{1\pm}=\pm \mathbf{t},\\
        &\mathbf{e}_{2\pm}=\pm\cos\frac{\phi}{2}\mathbf{n}- \sin\frac{\phi}{2}\mathbf{b},\\
        &\mathbf{e}_{3\pm}=\pm\sin\frac{\phi}{2}\mathbf{n}+ \cos\frac{\phi}{2}\mathbf{b},\\
        &\mathbf{l}_{\pm}=\eta_{\pm}\mathbf{e}_{1\pm}+\mathbf{e}_{2\pm},
\end{aligned}
\end{equation}
which governs the cross-sectional orientation of the two equivalent rods.
The Darboux frames are also shown in Fig. \ref{fig:sec24}(a).
According to Eq. \eqref{framedtof}, the angular velocity of the Darboux frame $\mathbf{\Omega}_\pm$ relates to the angular velocity of the Frenet frame $\mathbf{\Omega}_f$, following
\begin{equation}\label{avelo}
    \begin{aligned}
    \mathbf{\Omega}_\pm=\mathbf{\Omega}_f\pm\frac{\dot{\phi}}{2}\mathbf{t}.
    \end{aligned}
\end{equation}

The interactions between the two developable surfaces, including forces and moments, can be expressed as $\mathbf{F}_{inter}$ and $\mathbf{M}_{inter}$, which are shown in Fig. \ref{fig:sec24}(b). As stated in Eq. \eqref{ettfold}, the fold is modeled as a linear elastic torsional spring, leading to the result that the component of $\mathbf{M}_{inter}$ along $\mathbf{t}$ satisfies
\begin{equation}
    \mathbf{M}_{inter}\cdot\mathbf{t}=\frac{\partial V_{fold}}{\partial \phi}=k_c(2\arctan(\frac{\kappa_n}{\kappa_g})-\phi_0).
\end{equation}
Considering $\mathbf{M}_{inter}$ and $\mathbf{F}_{inter}$, the dynamical equations of rod$\pm$ from Eq. \eqref{eqdynamicfull} are
\begin{equation} \label{eqdynamicfull_fold}
    \begin{aligned}
         &\mathbf{F}_{\pm}^{\prime}\pm \mathbf{F}_{inter}=\rho h [(v_{0\pm}+\frac{1}{2}v_{0\pm}^2\gamma_{\pm})\ddot{\mathbf{r}}_0+(\frac{1}{2}v_{0\pm}^2+\frac{1}{3}v_{0\pm}^3\gamma_{\pm})(\dot{\bm{\Omega}}_{\pm}\times\mathbf{l}_{\pm}+\bm{\Omega}_{\pm}\times(\bm{\Omega}_{\pm}\times\mathbf{l}_{\pm}))\mp\frac{1}{3}v_{0\pm}^3\dot{\eta}_{\pm}\bm{\Omega}_{\pm}^{\prime}\times\mathbf{l}_{\pm}],\\
        &\mathbf{M}_{\pm}^{\prime}+\mathbf{r}^{\prime}_0\times \mathbf{F}_\pm\pm \mathbf{M}_{inter}=\rho h[(\frac{1}{2}v_{0\pm}^2+\frac{1}{3}v_{0\pm}^3\gamma_{\pm})\mathbf{l}_{\pm}\times\ddot{\mathbf{r}}_0+(\frac{1}{3}v_{0\pm}^3+\frac{1}{4}v_{0\pm}^4\gamma_{\pm})\mathbf{l}_{\pm}\times(\dot{\bm{\Omega}}_{\pm}\times\mathbf{l}_{\pm}+\bm{\Omega}_{\pm}\times(\bm{\Omega}_{\pm}\times\mathbf{l}_{\pm}))\\
        &\mp\frac{1}{4}v_{0\pm}^4\dot{\eta}_{\pm}\mathbf{l}_{\pm}\times(\bm{\Omega}_{\pm}^{\prime}\times\mathbf{l}_{\pm})],\\
        &\mathbf{M}_{\pm}=\pm[\frac{1}{\kappa_{n\pm}}\left(\frac{\d}{\d S} \frac{\partial \varepsilon_{\pm}}{\partial \eta_{\pm}^{\prime}}-\frac{\partial \varepsilon_{\pm}}{\partial \eta_{\pm}}\right) \mathbf{e}_{1\pm}+\left(-\frac{\partial \varepsilon_{\pm}}{\partial \kappa_n}-\frac{\eta_\pm}{\kappa_{n\pm}}(\frac{\d}{\d S} \frac{\partial \varepsilon_{\pm}}{\partial \eta_{\pm}^{\prime}}-\frac{\partial \varepsilon_{\pm}}{\partial \eta_{\pm}})\right) \mathbf{e}_{2\pm}+m_{3\pm} \mathbf{e}_{3\pm}],\\
        &\mathbf{M}_{inter}\cdot\mathbf{t}=\frac{\partial V_{fold}}{\partial \phi}=k_c(2\arctan(\frac{\kappa_n}{\kappa_g})-\phi_0),
    \end{aligned}
\end{equation}
where $\gamma_\pm=\pm\eta_\pm'\mp(1+\eta_\pm^2)\kappa_g$ and the boundary conditions are
\begin{equation}\label{eqbvp_de_fold}
    \begin{aligned}
        &\mathbf{F}_{\Sigma}=\mathbf{F}_e,\\
        &\mathbf{M}_{\Sigma}=\mathbf{M}_e.
    \end{aligned}
\end{equation}
Define $\mathbf{F}_\Sigma=\mathbf{F}_+ + \mathbf{F}_-$ and $\mathbf{M}_\Sigma=\mathbf{M}_++\mathbf{M}_-$. Eq. \eqref{eqdynamicfull_fold} can be simplified into
\begin{equation}\label{dynamic_sum}
    \begin{aligned}
        &\mathbf{F}'_\Sigma=\Sigma_\pm\rho h[(\frac{1}{2}v_{0\pm}^2+\frac{1}{3}v_{0\pm}^3\gamma_{\pm})\mathbf{l}_{\pm}\times\ddot{\mathbf{r}}_0+(\frac{1}{3}v_{0\pm}^3+\frac{1}{4}v_{0\pm}^4\gamma_{\pm})\mathbf{l}_{\pm}\times(\dot{\bm{\Omega}}_{\pm}\times\mathbf{l}_{\pm}+\bm{\Omega}_{\pm}\times(\bm{\Omega}_{\pm}\times\mathbf{l}_{\pm}))\mp\frac{1}{3}v_{0\pm}^3\dot{\eta}_{\pm}\bm{\Omega}_{\pm}^{\prime}\times\mathbf{l}_{\pm}],\\
        &\mathbf{M}'_\Sigma+\mathbf{t}\times\ \mathbf{F}'_\Sigma=\Sigma_\pm\rho h[(\frac{1}{2}v_{0\pm}^2+\frac{1}{3}v_{0\pm}^3\gamma_{\pm})\mathbf{l}_{\pm}\times\ddot{\mathbf{r}}_0+(\frac{1}{3}v_{0\pm}^3+\frac{1}{4}v_{0\pm}^4\gamma_{\pm})\mathbf{l}_{\pm}\times(\dot{\bm{\Omega}}_{\pm}\times\mathbf{l}_{\pm}+\bm{\Omega}_{\pm}\times(\bm{\Omega}_{\pm}\times\mathbf{l}_{\pm}))\\
        &\mp\frac{1}{4}v_{0\pm}^4\dot{\eta}_{\pm}\mathbf{l}_{\pm}\times(\bm{\Omega}_{\pm}^{\prime}\times\mathbf{l}_{\pm})],\\
        & \mathbf{M}_+\cdot\mathbf{t}+k_c(2\arctan(\frac{\kappa_n}{\kappa_g})-\phi_0)=\rho h[(\frac{1}{2}v_{0+}^2+\frac{1}{3}v_{0+}^3\gamma_{+})\mathbf{l}_{+}\times\ddot{\mathbf{r}}_0+(\frac{1}{3}v_{0+}^3+\frac{1}{4}v_{0+}^4\gamma_{+})\mathbf{l}_{+}\times(\dot{\bm{\Omega}}_{+}\times\mathbf{l}_{+}+\bm{\Omega}_{+}\times(\bm{\Omega}_{+}\times\mathbf{l}_{+}))\\
        &-\frac{1}{4}v_{0+}^4\dot{\eta}_{+}\mathbf{l}_{+}\times(\bm{\Omega}_{+}^{\prime}\times\mathbf{l}_{+})]\cdot\mathbf{t},\\
        &\mathbf{M}_{\pm}=\pm[\frac{1}{\kappa_{n\pm}}\left(\frac{\d}{\d S} \frac{\partial \varepsilon_{\pm}}{\partial \eta_{\pm}^{\prime}}-\frac{\partial \varepsilon_{\pm}}{\partial \eta_{\pm}}\right) \mathbf{e}_{1\pm}+\left(-\frac{\partial \varepsilon_{\pm}}{\partial \kappa_{n\pm}}-\frac{\eta_\pm}{\kappa_{n\pm}}(\frac{\d}{\d S} \frac{\partial \varepsilon_{\pm}}{\partial \eta_{\pm}^{\prime}}-\frac{\partial \varepsilon_{\pm}}{\partial \eta_{\pm}})\right) \mathbf{e}_{2\pm}+m_{3\pm} \mathbf{e}_{3\pm}].
    \end{aligned}
\end{equation}
 The variables in Eq. \eqref{dynamic_sum} are $\mathbf{r}_0,\mathbf{F}_\Sigma,m_{3\pm}$, thus the simplification can avoid solving $\mathbf{F}_{inter},\mathbf{F}_\pm$ and $\mathbf{M}_{inter}$. Numerically solving Eq. \eqref{dynamic_sum} requires expressing the governing equations in component forms within a fixed frame. The detailed steps are illustrated in \ref{AppB}.

In this section, we demonstrate the dynamical theory of curve-fold origami and develop a numerically solvable bi-rod model. Illustrative examples of dynamic developable surfaces and curve-fold origami are demonstrated in the following section within the framework established in Sections \ref{sec2} and \ref{sec3new}.

\section{Illustrative examples} \label{sec3}
To validate the dynamical theories established in this work, we demonstrate three examples of dynamic developable surfaces and curve-fold origami in this section. Theoretical results are presented and compared with FEM (finite element method) results. The finite element results of the examples are obtained using the software Abaqus, where developable surfaces are modeled as S4R shell elements and folds are modeled as arrays of join-rotation connectors\citep{WEN2024105829,FEM,water}.

\subsection{Dynamics of a curve-folded plate} \label{sec3.2}

\begin{figure}[!t]
    \centering
    \includegraphics[width=1\textwidth]{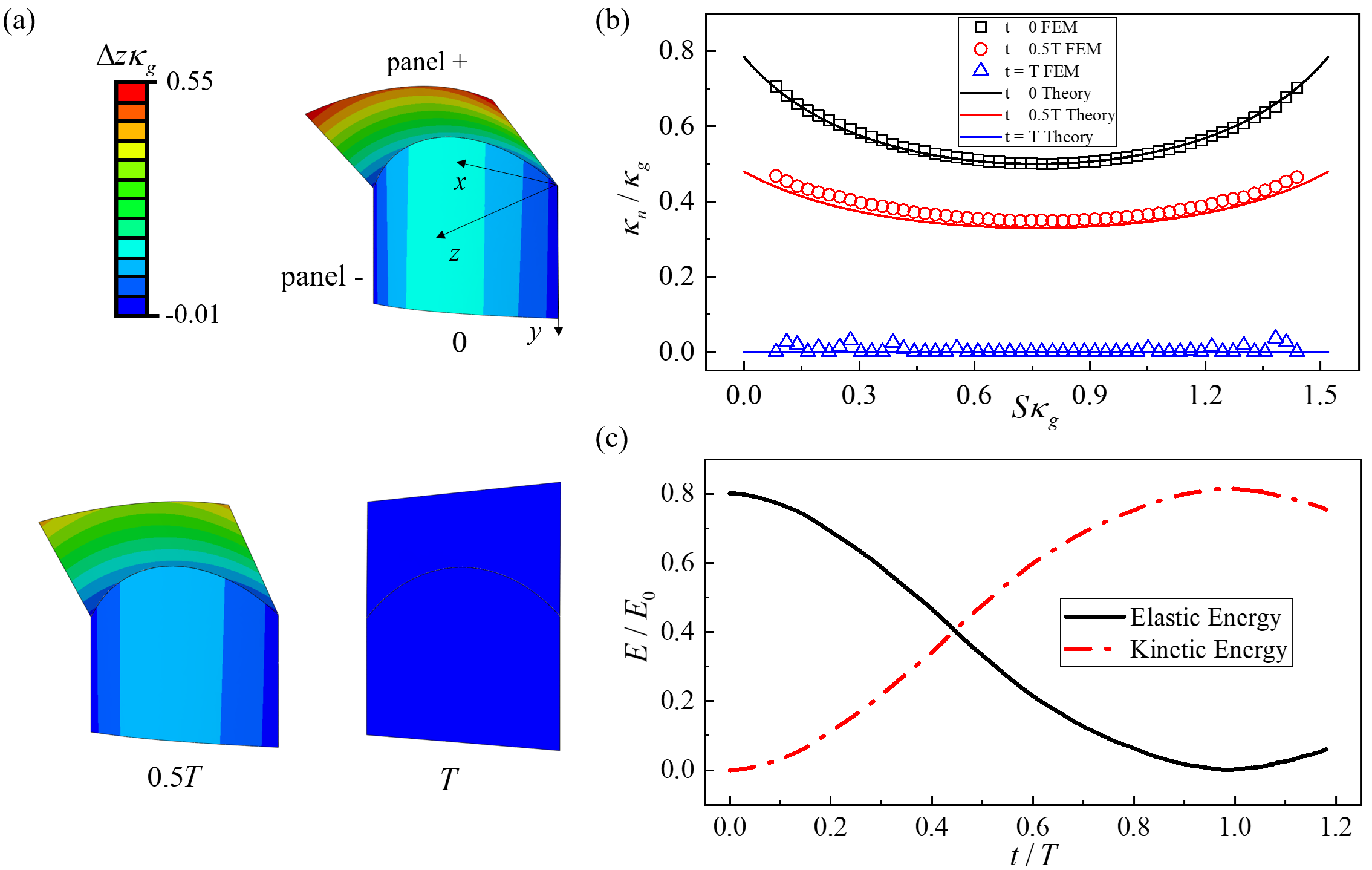}
    \caption{\label{fig:sec42} (a) FEM result of the dynamics of a curve-folded plate. (b) Comparison between the theoretical results and FEM results. (c) FEM results of the evolution of kinetic and elastic energy. $T$ is the characteristic time shown in (a). \comment{$\kappa_g$ is the initial geodesic curvature of the crease, and $E_0$ is defined as $E_0=0.5D\kappa_g^2Lb$ to nondimensionalize the energy.}}
\end{figure}

A curve-folded plate is a basic cell of curve-fold origami. In this example, we analyse a freely developing curved-folded plate (length: $L$, width: $b$, thickness: $h$) where the fold starts at the middle point along the length. The specific loading process is to press the two long edges to fold the plate statically, and then release the compression to unfold it dynamically. Previous studies indicate that the generators remain parallel in the static deformation \citep{lee2018elastica}. Here, we first assume that this conclusion remains applicable in dynamics and will verify it subsequently. Under such an assumption, the geometric variants are
\begin{equation}\label{evetaplate}
\begin{aligned}
        &\eta_{\pm}=\tan(\frac{2S-S_0}{2}\kappa_g),\\
        &v_{0\pm}=\cos(\frac{2S-S_0}{2}\kappa_g)\frac{L}{2}\mp\frac{\cos(\frac{2S-S_0}{2}\kappa_g)}{\kappa_g}(\cos(\frac{2S-S_0}{2}\kappa_g)-\cos(\frac{S_0}{2}\kappa_g)),\\
        &\gamma_{\pm}=\pm\eta_{\pm}^{\prime}\mp(1+\eta_{\pm}^2)\kappa_g=0.
\end{aligned}
\end{equation}
Eq. \eqref{eqdynamicfull_fold} is thus simplified into
\begin{equation} \label{eqdynamicfull_fold_eplate}
    \begin{aligned}
         &\mathbf{F}_{\pm}^{\prime}\pm \mathbf{F}_{inter}=\rho h [v_{0\pm}\ddot{\mathbf{r}}_0+\frac{1}{2}v_{0\pm}^2(\dot{\bm{\Omega}}_{\pm}\times\mathbf{l}_{\pm}+\bm{\Omega}_{\pm}\times(\bm{\Omega}_{\pm}\times\mathbf{l}_{\pm}))],\\
        &\mathbf{M}^{\prime}+\mathbf{r}^{\prime}_0\times \mathbf{F}\pm \mathbf{M}_{inter}=\rho h[\frac{1}{2}v_{0\pm}^2\mathbf{l}_{\pm}\times\ddot{\mathbf{r}}_0+\frac{1}{3}v_{0\pm}^3\mathbf{l}_{\pm}\times(\dot{\bm{\Omega}}_{\pm}\times\mathbf{l}_{\pm}+\bm{\Omega}_{\pm}\times(\bm{\Omega}_{\pm}\times\mathbf{l}_{\pm})),\\
        &\mathbf{M}_\pm=Dv_0(\mp2(1+\eta_\pm^4) \eta_\pm\kappa_{n\pm}+\frac{1}{2}v_0((1+\eta_\pm^2)\kappa_{n\pm})')\mathbf{e}_1- Dv_0(\pm\kappa_{n\pm}(1-\eta_\pm^2)+\frac{\eta_\pm}{2}v_0((1+\eta_\pm^2)\kappa_{n\pm})')\mathbf{e}_2+m_3\mathbf{e}_3,\\
        &\mathbf{M}_{inter}\cdot\mathbf{t}=\frac{\partial V_{fold}}{\partial \phi}=k_c(2\arctan(\frac{\kappa_n}{\kappa_g})-\phi_0),
    \end{aligned}
\end{equation}
where $\eta_{\pm}$ and $v_{0\pm}$ are given in Eq. \eqref{evetaplate}. Solving Eq. \eqref{eqdynamicfull_fold_eplate} yields $\mathbf{\Omega}_-\times\mathbf{l-}=0$, indicating that the solution does not violate the assumption that the generators remain parallel and the panels deform cylindrically. Since the dynamic process is uniquely determined by Eq. \eqref{eqdynamicfull_fold}, the solution of Eq. \eqref{eqdynamicfull_fold_eplate} is the unique solution that satisfies Eq. \eqref{eqdynamicfull} in this example. The process is also simulated in finite element analysis, where the mechanical parameters are selected as $D\kappa_g/k_c=1$, and the geometric parameters are $L/b=1/2$. The good agreement between the theoretical and numerical results is demonstrated in Fig. \ref{fig:sec42}(b). Fig. \ref{fig:sec42}(c) illustrates the conversion between elastic energy and kinetic energy, which further demonstrates that the elastic energy is completely converted into kinetic energy and the origami structure is fully flattened at time $T$. $E_0$ in Fig. \ref{fig:sec42} is used to nondimensionalize the energy, defined as $E_0=0.5D\kappa_g^2Lb$.

\subsection{Dynamics of single-vertex multi-curve-fold origami} \label{sec3.3}

{Our previous work \citep{WEN2024105829} demonstrates that the deformed configuration of a single-vertex origami is determined by the three geometric constraints, which are the neighboring folds constraints}
\begin{equation} \label{eq049}
\left(1+\eta_{2-}^2\left(S_2\right)\right) \kappa_{n2-}\left(S_2\right)=\frac{\left(1+\eta_{1+}^2\left(S_1\right)\right) \kappa_{n1+}\left(S_1\right)}{1+v_{0+}\left(S_1\right) \eta_{1+}^{\prime}\left(S_1\right)-v_{0+}\left(S_1\right) \kappa_{g1+}\left(S_1\right)-v_{0+}\left(S_1\right) \eta_{1+}^2\left(S_1\right) \kappa_{g1+}\left(S_1\right)},
\end{equation}
\begin{equation}\label{eq050}
\tau_2(S_2) = -\eta_{2-}(S_2) \kappa_{n2-}(S_2)-\frac{\kappa_{g2-}(S_2)}{\kappa_{2}^2(S_2)} \kappa_{n2-}^{\prime}(S_2)+\frac{\kappa_{n2-}(S_2)}{\kappa_{2}^2(S_2)} \kappa_{g2-}^{\prime}(S_2),
\end{equation}
periodic constraints 
\begin{equation} \label{eqperi}
    \begin{aligned}
    \kappa_{i+2}(S)&=\kappa_i(S),\\
    \tau_{i+2}(S)&=\tau_i(S),
    \end{aligned}
\end{equation}
and vertex constraints
\begin{equation} \label{eq055}
\begin{aligned}
    [(\sin \frac{\phi_1(0)}{2}\sin\frac{\phi_2(0)}{2})^{-1}-\cot \frac{\phi_1(0)}{2} \cot\frac{\phi_2(0)}{2}]\cos \alpha&=1. \\
\end{aligned}
\end{equation}

\begin{figure}[!h]
    \centering
    \includegraphics[width=1\textwidth]{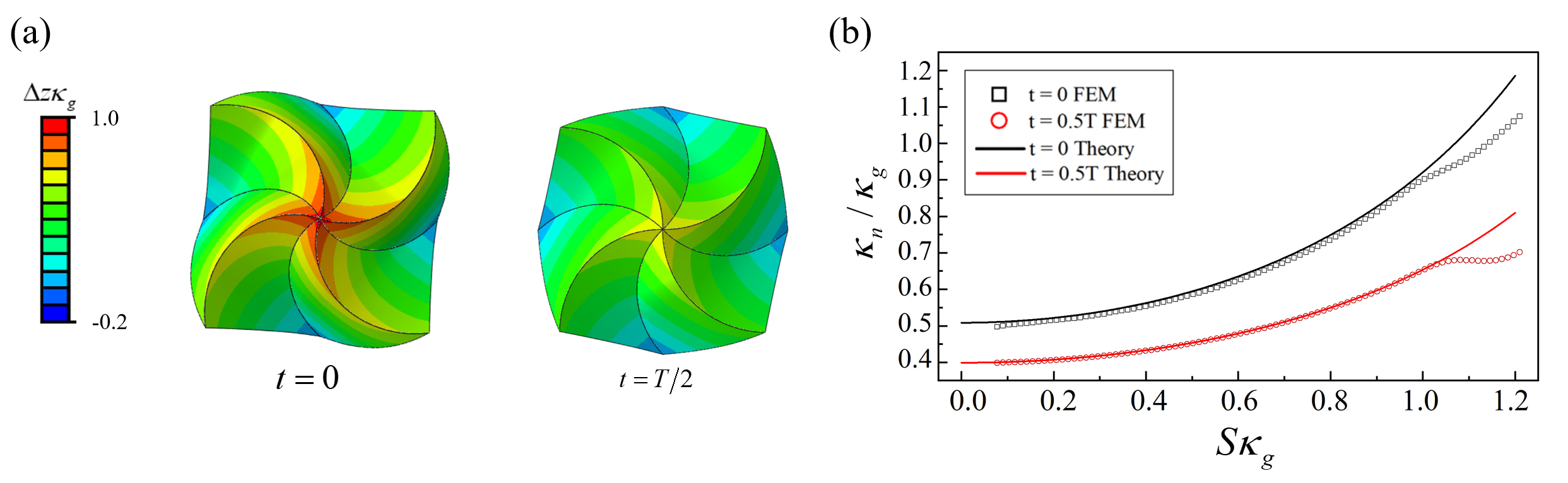}
    \caption{\label{fig:vertex} (a) FEM result of the dynamics of the single-vertex curved origami. (b) Comparison between the theoretical results and FEM results.  $T$ is the characteristic time between the initial state and the flat state. \comment{$\kappa_g$ is the geodesic curvature of the crease, which is used to nondimensionalize the curvature.}}
\end{figure}
The constraints strictly limit the configuration space of the folds, resulting in the deformation of most of the fold being determined by the vertex folding angle in static deformation, and is thus approximated as
\begin{equation} \label{e056}
\begin{gathered}
\kappa\left(S\right)=\sum_{i=0}^n \frac{1}{i !} \kappa^{(i)}\left(0\right) S^i, \\
\tau\left(S\right)=\sum_{i=0}^n \frac{1}{i !} \tau^{(i)}\left(0\right) S^i.
\end{gathered}
\end{equation}
While the proof is purely geometric and does not introduce any mechanical variables, the conclusion can be extended to dynamical situations.  {Given the rotation angle at the vertex, the geometry of the folds can be calculated, which is independent of the elasticity according to the result in the static study. } The comparison between the FEM results and the theoretical results derived from Eq. \eqref{e056} is shown in Fig. \ref{fig:vertex}. The good agreement validates the conclusion in dynamical situations.

\subsection{Dynamics of a cut-then-fold ring origami}
\begin{figure}[!b]
    \centering
    \includegraphics[width=1\textwidth]{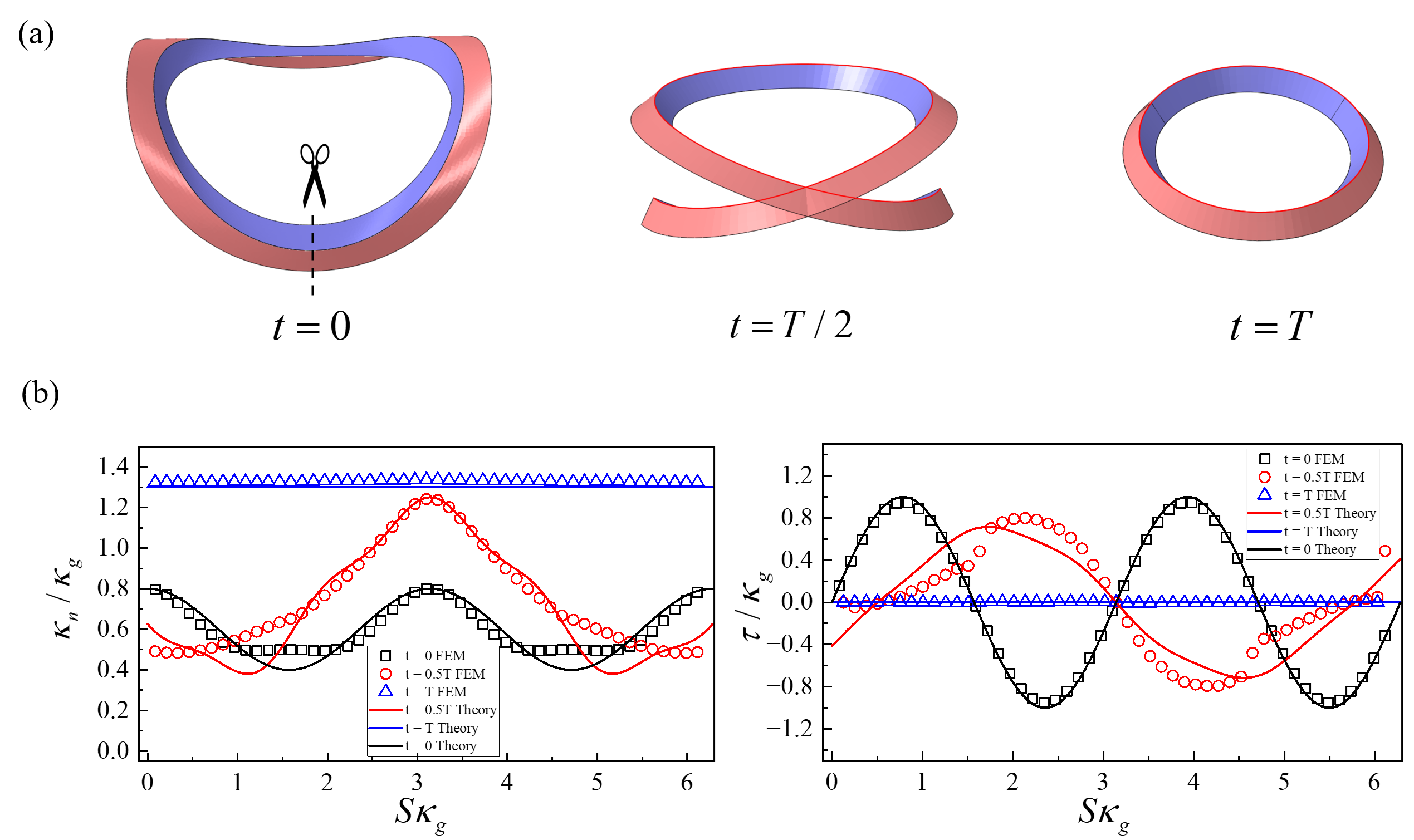}
    \caption{\label{fig:sec52} (a) Finite element result of the dynamics of a cut-then-folded ring. (b) Comparison between the theoretical results and FEM results. $T$ is the characteristic time shown in (a). \comment{$\kappa_g$ denotes the initial geodesic curvature of the crease, which is used to nondimensionalize the curvature and torsion.} } 
\end{figure}

The equilibrium configurations of ring origami, in both closed and open forms, have been analyzed in \citet{dias2012geometric,DIAS201457}. These studies show that a folded closed ring buckles out of plane and acquires nonzero torsion along the crease, whereas a folded open ring remains planar with zero torsion. In our dynamic example, we examine the evolution of a cut folded closed ring. The resulting motion gradually drives the structure from an initial out-of-plane configuration to a planar configuration, as illustrated in Fig.~\ref{fig:sec52}.

We consider a circular ring with small width, i.e., $v_0\kappa_g\to0$. Eq. \eqref{eqdynamicfull_fold} is simplified into
\begin{equation} \label{eqdynamicfull_fold_thin}
    \begin{aligned}
         &\mathbf{F}^{\prime}\pm \mathbf{F}_{inter}=\rho h v_0\ddot{\mathbf{r}}_0,\\
        &\mathbf{M}^{\prime}+\mathbf{r}^{\prime}_0\times \mathbf{F}\pm \mathbf{M}_{inter}=\frac{1}{2}\rho h v_0^2\mathbf{l}\times\ddot{\mathbf{r}}_0+\frac{1}{3}v_0^3\mathbf{l}\times(\dot{\bm{\Omega}}\times\mathbf{l}+\bm{\Omega}\times(\bm{\Omega}\times\mathbf{l})),\\
        &\mathbf{M}=-2(1+\eta^4)Dv_0 \kappa_n\mathbf{e}_1-Dv_0 \kappa_n(1-\eta^2)\mathbf{e}_2+m_3\mathbf{e}_3,\\
        &\mathbf{M}_{inter}\cdot\mathbf{t}=\frac{\partial V_{fold}}{\partial \phi}=k_c(2\arctan(\frac{\kappa_n}{\kappa_g})-\phi_0).
    \end{aligned}
\end{equation}
For a smooth closed ring, the boundary conditions are
\begin{equation}
    \begin{aligned}
        &\kappa^{(n)}(S_0)=\kappa^{(n)}(0),\\
        &\tau^{(n)}(S_0)=\tau^{(n)}(0),
    \end{aligned}
\end{equation}
and the boundary values in Eq. \eqref{eqvari1} automatically equal zero. After the cut, the closure constraint ($\mathbf{r}_0(0)=\mathbf{r}_0(S_0)$) fails, and the boundary conditions become $\mathbf{F}=0$ and $\mathbf{M}=0$.

Numerically solving Eq.~\eqref{eqdynamicfull_fold_thin} yields the curvature and torsion distributions at any time $t$. The theory indicates that {the torsion of the cut ring gradually decreases while the curvature increases.} This behavior is reproduced in finite-element simulations, as shown in Fig.~\ref{fig:sec52}, where the mechanical parameters satisfy $D\kappa_g/k_c=1$ and $\phi_0=\pi$. {
A direct comparison between the theoretically predicted and numerically simulated curvature and torsion is presented in Fig. \ref{fig:sec52}, showing reasonably good quantitative agreement overall. Nevertheless, noticeable discrepancies persist in this example. We attribute this primarily to the numerical difficulty of handling the abrupt geometric boundary change induced by the initial cut. In addition, the assumption underlying Eq. (\ref{eqdynamicfull_fold_thin}) may break down even in the narrow-ribbon limit \citep{starostin2007shape}.
}

\section{Discussion} \label{sec:conclusion}
In this work, we have established an exact dimension-reduced dynamic theory for developable surfaces, which is extended to the realm of curve-fold origami with wide panels. We exploit the 1D nature of developable surfaces under isometric constraint and
use the Wunderlich functional to derive the analytical dynamic equations for a reference curve, similar to the dynamic rod equations, with accurate predictions for the dynamics of developable surfaces. The theoretical results predicted by our model and the FEM results agree quantitatively well, validating the accuracy and versatility of our dynamic theory.

The key contributions of this study can be summarized as follows:

(1) By considering the variation of local frames, we have 
established an exact Lagrangian (including the elastic and kinetic energies) of developable surfaces under isometric constraint, which leads to the rod-like dynamic equations of a reference curve.
An equivalent exact nonlinear dynamic rod model is developed to describe the dynamic behaviors of the reference curve and the developable surface accurately.

(2) By employing a pair of developable surfaces connected by a crease, we have formulated a dynamic theory for curve-fold origami with wide panels. This theory is represented by an equivalent nonlinear bi-rod model, which captures the dynamic behavior of curve-fold origami structures accurately.

(3) To validate our theoretical framework, we have solved the dynamic evolution of developable surfaces and curve-fold origami in four distinct scenarios. The results obtained from our theoretical models are compared with those from FEM, demonstrating the accuracy and versatility of our proposed theories.

 We believe that our dynamical theory for developable surfaces and curved origami, including the equivalent rod models, can deepen understanding of the physics of the curve-fold origami system and provide new insights into shape programming. 
 {Our theoretical framework can be naturally extended to developable surfaces with reinforcing ribs, such as those in space antennas. For example, we may optimize a developable curved surface as part of an antenna. Within our dynamic framework, any admissible reference curve can be selected to parameterize and describe the motion of a developable surface. By choosing the reinforced rib as the reference curve, the dynamics of the rib can be naturally incorporated into the governing equations of our model, leading to a unified set of dynamic equations for a rib-reinforced developable surface. Since the present dynamic model for developable surfaces and the classical Kirchhoff rod model are both formulated as one-dimensional rod theories, their coupling is conceptually straightforward and mathematically consistent. Another important direction for future work is to extend the present dynamic framework to developable surfaces with intersecting generators and more complex generator distributions. For instance, a closed Möbius ring necessarily involves intersections of generators \citep{starostin2007shape}, leading to geometric singularities that require a refined treatment. Moreover, dynamic bifurcation phenomena, similar to those reported in \citep{wen2025geometry}, are expected to emerge in these more intricate configurations.}
 We anticipate that our work can contribute to the design of complex dynamic curve-fold origami structures and developable surfaces in the fields of robotics, metamaterials, and architecture.
\section*{CRediT authorship contribution statement}
\textbf{Zhixuan Wen:} Conceptualization, Methodology, Software, Validation, Formal analysis, Writing – original draft. \textbf{Sheng Mao:} Methodology, Writing– review \& editing.  \textbf{Huiling Duan:} Conceptualization, Methodology, Writing – review \& editing, Supervision, Funding acquisition.  \textbf{Fan Feng:} Conceptualization, Investigation, Methodology, Software, Writing– review \& editing, Project administration, Funding acquisition.
\section*{Declaration of competing interest}
The authors declare that they have no known competing financial interests or personal relationships that could have appeared to influence the work reported in this paper.
\section*{Data availability}
No data was used for the research described in the article.
\section*{Acknowledgement}
Financial support for this work was provided by the National Natural Science Foundation of China (grant nos. 11988102, 12472061) and the National Key Research and Development Program of China (grant no. 2020YFE0204200).

\appendix

\setcounter{table}{0}   
\setcounter{figure}{0}
\setcounter{equation}{0}

\renewcommand\thetable{A.\arabic{table}}
\renewcommand\thefigure{A.\arabic{figure}}
\renewcommand\theequation{A.\arabic{equation}}
\section{Dynamical equation in the Eulerian framework}\label{AppA}
In this appendix, we demonstrate that the equations derived from the Eulerian and Lagrangian descriptions are equivalent.

\begin{figure}[!b]
    \centering
    \includegraphics[width=1\textwidth]{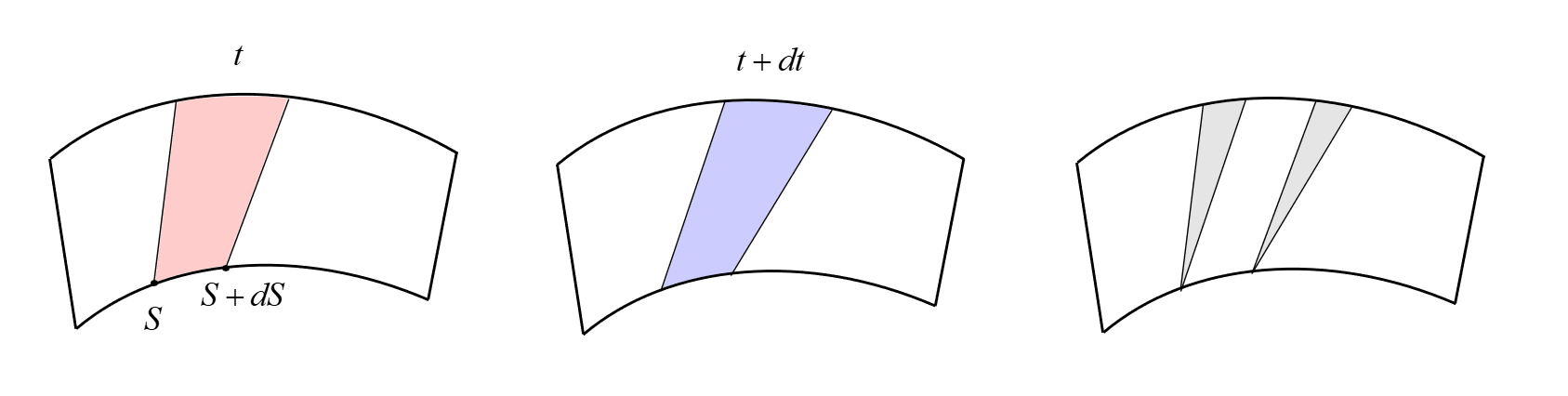}
    \caption{\label{app} Illustration of the regions associated with linear and angular momentum at times  $t$ and $t+\d t$}
\end{figure}

In the Eulerian framework, the dynamical equation is
\begin{equation} \label{eqdynamica1}
    \begin{aligned}
        &\mathbf{F}^{\prime}+\dot{\mathbf{p}}_{in}^*=\dot{\mathbf{p}}^*,\\
        &\mathbf{M}+\mathbf{r}^{\prime}_0\times \mathbf{F}+\dot{\mathbf{p}}_{\theta in}^*=\dot{\mathbf{p}}_{\theta}^*,\\
        &\mathbf{M}=\frac{1}{\kappa_n}\left(\frac{\d}{\d S} \frac{\partial \varepsilon}{\partial \eta^{\prime}}-\frac{\partial \varepsilon}{\partial \eta}\right) \mathbf{e}_1+\left(-\frac{\partial \varepsilon}{\partial \kappa_n}-\frac{\eta}{\kappa_n}(\frac{\d}{\d S} \frac{\partial \varepsilon}{\partial \eta^{\prime}}-\frac{\partial \varepsilon}{\partial \eta})\right) \mathbf{e}_2+m_3 \mathbf{e}_3,
    \end{aligned}
\end{equation}
where $\mathbf{p}^*$ and $\mathbf{p}_{\theta}^*$ are the unit-length momentum and momentum of the region $(S,S+\d S)$. The illustration of the region at the time $t$ and $t+\d t$ is shown in Fig. \ref{app}. Notice that the mass in the region $(S,S+\d S)$ varies over time. The net momentum and angular momentum input $\dot{\mathbf{p}}_{in}^*$ and $\dot{\mathbf{p}}_{in \theta}^*$ are considered, which is introduced by the gray region shown in Fig. \ref{app}.
Therefore, the derivatives of $\mathbf{p}^*$ and $\mathbf{p}_{\theta}^*$ are
\begin{equation}
    \begin{aligned}
        \dot{\mathbf{p}}^*&=\rho h(\int_0^{v_0}\partial_t(\left|\mathbf{J}\right|\mathbf{u})\d v+\left|\mathbf{J}(v_0)\right|\mathbf{u}(v_0)\dot{v}_0)\\
        &=\rho h (\dot{\eta}'\int_0^{v_0}\mathbf{u}v\d v-2\eta\dot{\eta}\int_0^{v_0}\mathbf{u}v\d v+\int_0^{v_0}\left|\mathbf{J}\right|(\ddot{\mathbf{r}}_0+\dot{\mathbf{\Omega}}\times\mathbf{l}+\mathbf{\Omega}\times(\mathbf{\Omega}\times\mathbf{l})+\dot{\eta}\mathbf{\Omega}\times\mathbf{e}_1)\d v+\left|\mathbf{J}(v_0)\right|\mathbf{u}(v_0)\dot{v}_0),\\
        \dot{\mathbf{p}}_{\theta}^*&=\rho h(\int_0^{v_0}v\mathbf{l}\times\partial_t(\left|\mathbf{J}\right|\mathbf{u})\d v+\int_0^{v_0}\left|\mathbf{J}\right|(\dot{\mathbf{r}}_0+v\dot{\mathbf{l}})\times\mathbf{u}\d v
        +\left|\mathbf{J}(v_0)\right|v_0\mathbf{l}(v_0)\times\mathbf{u}(v_0)\dot{v}_0)\\
        &=\rho h (\dot{\eta}'\int_0^{v_0}v^2\mathbf{l}\times\mathbf{u}\d v-2\eta\dot{\eta}\int_0^{v_0}v^2\mathbf{l}\times\mathbf{u}\d v+\int_0^{v_0}\left|\mathbf{J}\right|v\mathbf{l}\times(\ddot{\mathbf{r}}_0+\dot{\mathbf{\Omega}}\times\mathbf{l}+\mathbf{\Omega}\times(\mathbf{\Omega}\times\mathbf{l})+\dot{\eta}\mathbf{\Omega}\times\mathbf{e}_1)\d v\\
&+\int_0^{v_0}\left|\mathbf{J}\right|\dot{\eta}\mathbf{e}_1\times\mathbf{u}\d v+\left|\mathbf{J}(v_0)\right|v_0\mathbf{l}(v_0)\times\mathbf{u}(v_0)\dot{v}_0),
    \end{aligned}
\end{equation}
where the reference point of the angular momentum is a fixed point with the same coordinate as $\mathbf{r}_0(S,t)$. The net momentum and angular momentum input $\dot{\mathbf{p}}_{in}^*$ and $\dot{\mathbf{p}}_{in \theta}^*$ equal the momentum difference in the gray region
\begin{equation}
    \begin{aligned}
        \dot{\mathbf{p}}_{in}^*&=\partial_S(\rho h \dot{\eta}\int_0^{v_0}\mathbf{u} v \d v)=\rho h (\dot{\eta}'\int_0^{v_0}\mathbf{u} v \d v+\dot{\eta}\int_0^{v_0}\mathbf{\Omega}\times(\eta v\mathbf{l}+(1+\eta'v-(1+\eta^2)\kappa_g v)\mathbf{e}_1 )v \d v+\dot{\eta}\mathbf{u}(v_0)v_0v_0'),\\
        \dot{\mathbf{p}}_{\theta in}^*&=\rho h (\dot{\eta}\int_0^{v_0}v^2\mathbf{l}\times\mathbf{u}' \d v
     +\dot{\eta}'\int_0^{v_0}v^2\mathbf{l}\times\mathbf{u} \d v
     + \dot{\eta}\int_0^{v_0}v(\mathbf{r}_0'+v\mathbf{l}')\times \mathbf{u}\d v)\\
     &=\rho h (\dot{\eta}'\int_0^{v_0}v^2\mathbf{l}\times\mathbf{u}  \d v+\dot{\eta}\int_0^{v_0}v^2\mathbf{l}\times(\mathbf{\Omega}\times(\eta v\mathbf{l}+(1+\eta'v-(1+\eta^2)\kappa_g v)\mathbf{e}_1 )) \d v\\
     &+\dot{\eta}\int_0^{v_0}v(\eta v\mathbf{l}+(1+\eta'v-(1+\eta^2)\kappa_g v)\mathbf{e}_1)\times\mathbf{u} \d v +\dot{\eta}\mathbf{l}(v_0)\times\mathbf{u}(v_0)).
    \end{aligned}
\end{equation}
\begin{figure}[t!]
    \centering
    \includegraphics[width=0.6\textwidth]{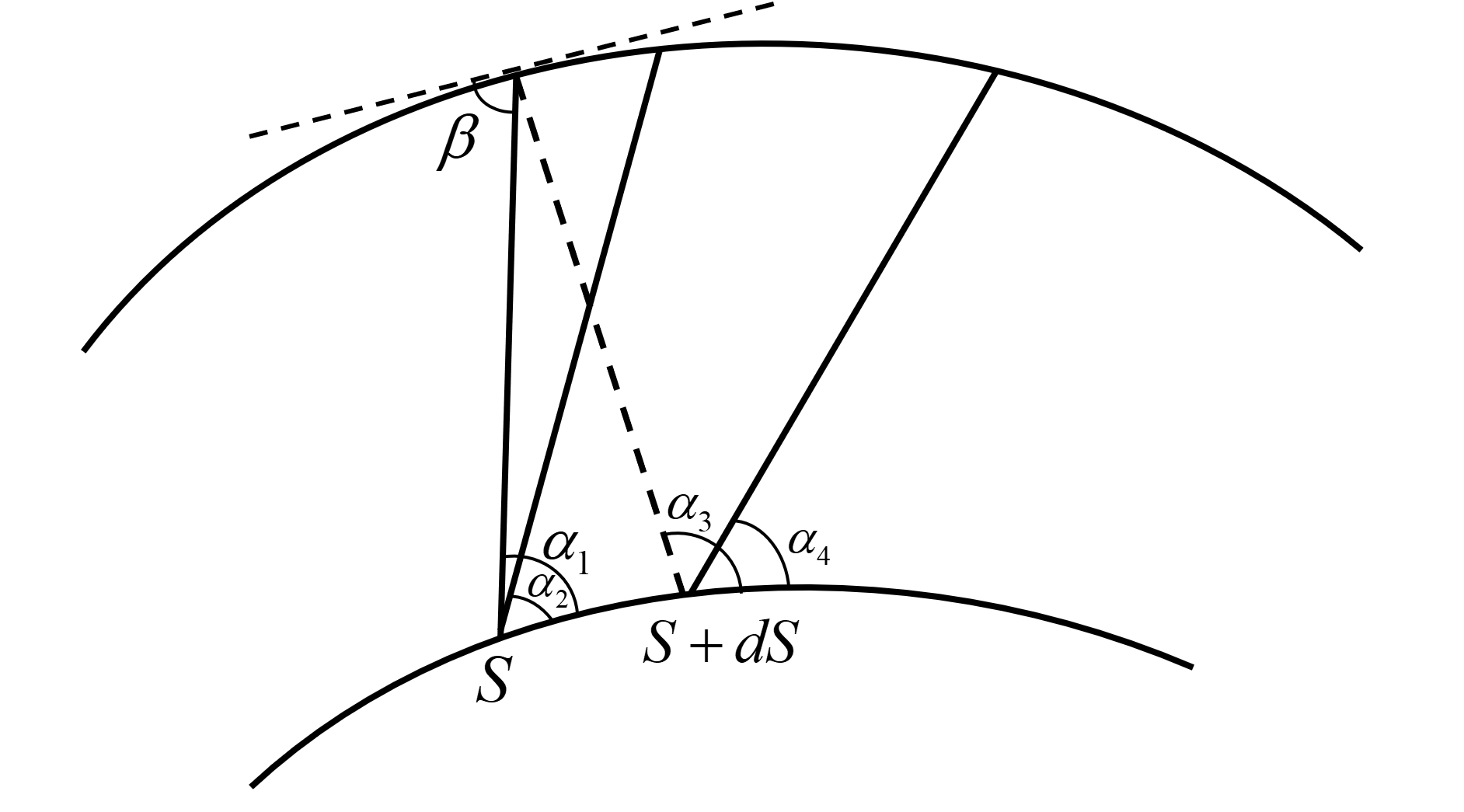}
    \caption{\label{app2} Illustrations for deriving the geometric correlations between $v_0^{\prime}$ and $\dot{v}_0$.}
\end{figure}

Here $\dot{v}_0$ can be expressed in terms of $v_0^\prime$, $\eta^\prime$ and $\dot{\eta}$ in the following derivation. As shown in Fig. \ref{app2}, the geometric correlations exist as follows
\begin{equation}
    \begin{aligned}
    &\eta_i=\cot \alpha_i,\\
        &\eta_2-\eta_1=\dot{\eta} \d t,\\
        &\eta_4-\eta_1=\eta^{\prime} \d S,\\
        &(\eta_3-\eta_1)v_{01}=((1+\eta_1^2)\kappa_g v_{01}-1)\d S,\\
        &v_{02}-v_{01}=v_{01}(\cot \beta-\eta_1)(\alpha_1-\alpha_2),\\
        &v_{04}-v_{03}=v_{03}(\cot \beta-\eta_3)(\alpha_3-\alpha_4),\\
        &v_{03}-v_{01}=-\frac{\eta_1}{1+\eta_1^2}\d S+v_0\eta_1(\alpha_3-\alpha_1),\\
        &\dot{v}_0=\frac{1}{\d t}(v_{02}-v_{01}),\\
        &v^{\prime}_0=\frac{1}{\d S}(v_{04}-v_{01}).
    \end{aligned}
\end{equation}
Through a series of derivations, we derive the correlations between $\dot{v}_0$ and $v'_0$ as 
\begin{equation}
    \dot{v}_0=\frac{v_0'v_0\dot{\eta}+\kappa_g v_0^2\eta\dot{\eta}}{1+\eta'v_0-(1+\eta^2)\kappa_gv_0},
\end{equation}
and the specific form of the dynamical equation is thus derived
\begin{equation} \label{eqdynamicfullA1}
    \begin{aligned}
         &\mathbf{F}^{\prime}=\rho h [(v_0+\frac{1}{2}v_0^2\gamma)\ddot{\mathbf{r}}_0+(\frac{1}{2}v_0^2+\frac{1}{3}v_0^3\gamma)(\dot{\bm{\Omega}}\times\mathbf{l}+\bm{\Omega}\times(\bm{\Omega}\times\mathbf{l}))-\frac{1}{3}v_0^3\dot{\eta}\bm{\Omega}^{\prime}\times\mathbf{l}],\\
        &\mathbf{M}+\mathbf{r}^{\prime}_0\times \mathbf{F}=\rho h[(\frac{1}{2}v_0^2+\frac{1}{3}v_0^3\gamma)\mathbf{l}\times\ddot{\mathbf{r}}_0+(\frac{1}{3}v_0^3+\frac{1}{4}v_0^4\gamma)\mathbf{l}\times(\dot{\bm{\Omega}}\times\mathbf{l}+\bm{\Omega}\times(\bm{\Omega}\times\mathbf{l}))-\frac{1}{4}v_0^4\dot{\eta}\mathbf{l}\times(\bm{\Omega}^{\prime}\times\mathbf{l})],\\
        &\mathbf{M}=\frac{1}{\kappa_n}\left(\frac{\d}{\d S} \frac{\partial \varepsilon}{\partial \eta^{\prime}}-\frac{\partial \varepsilon}{\partial \eta}\right) \mathbf{e}_1+\left(-\frac{\partial \varepsilon}{\partial \kappa_n}-\frac{\eta}{\kappa_n}(\frac{\d}{\d S} \frac{\partial \varepsilon}{\partial \eta^{\prime}}-\frac{\partial \varepsilon}{\partial \eta})\right) \mathbf{e}_2+m_3 \mathbf{e}_3,
    \end{aligned}
\end{equation}
which is the same as Eq. \eqref{eqdynamicfull}. 
\newpage
\renewcommand\thetable{B.\arabic{table}}
\renewcommand\thefigure{B.\arabic{figure}}
\renewcommand\theequation{B.\arabic{equation}}

\section{Component form of the dynamical equations in the fixed frame}\label{AppB}
{
In this appendix, we transform the dynamical equations into component forms expressed in the $xyz$ frame, for the convenience of numerical solution. An algorithmic flow description is summarized in the end, and the corresponding equations are recalled.}

The first step is to unify the frames of rod $\pm$ in Eq. \eqref{eqdynamicfull_fold} into the Frenet frame. Substituting Eq. \eqref{framedtof} into Eq. \eqref{eqdynamicfull_fold}, we have

\begin{equation}\label{B1}
    \begin{aligned}
&\mathbf{F}_{\Sigma}^{\prime}=\Sigma_\pm \rho h \bigg[ \left(v_{0\pm} + \frac{1}{2}v_{0\pm}^2 \gamma_{\pm}\right) \ddot{\mathbf{r}}_0 + \left(\frac{1}{2}v_{0\pm}^2 + \frac{1}{3}v_{0\pm}^3 \gamma_{\pm}\right) \bigg( \dot{\bm{\Omega}}_{\pm} \times \left(\pm \eta_{\pm} \mathbf{t} \pm\cos\frac{\phi}{2}\mathbf{n}- \sin\frac{\phi}{2}\mathbf{b}\right)\\ &\quad\quad\quad\quad\quad\quad+ \bm{\Omega}_{\pm} \times \left(\bm{\Omega}_{\pm} \times \left(\pm \eta_{\pm} \mathbf{t} \pm\cos\frac{\phi}{2}\mathbf{n}- \sin\frac{\phi}{2}\mathbf{b}\right)\right) \bigg) \mp \frac{1}{3}v_{0\pm}^3 \dot{\eta}_{\pm} \bm{\Omega}_{\pm}^{\prime} \times \left(\pm \eta_{\pm} \mathbf{t} \pm\cos\frac{\phi}{2}\mathbf{n}- \sin\frac{\phi}{2}\mathbf{b}\right) \bigg], \\
&\mathbf{M}_{\Sigma}^{\prime} + \mathbf{r}_0^{\prime} \times \mathbf{F}_\Sigma= \Sigma_\pm \rho h \bigg[ \left(\frac{1}{2}v_{0\pm}^2 + \frac{1}{3}v_{0\pm}^3 \gamma_{\pm}\right) \left(\pm \eta_{\pm} \mathbf{t} \pm\cos\frac{\phi}{2}\mathbf{n}- \sin\frac{\phi}{2}\mathbf{b}\right) \times \ddot{\mathbf{r}}_0  \\
&\quad\quad\quad\quad\quad\quad\quad\quad\quad\quad+\left(\frac{1}{3}v_{0\pm}^3 + \frac{1}{4}v_{0\pm}^4 \gamma_{\pm}\right) \left(\pm \eta_{\pm} \mathbf{t} \pm\cos\frac{\phi}{2}\mathbf{n}- \sin\frac{\phi}{2}\mathbf{b}\right) \times \bigg( \dot{\bm{\Omega}}_{\pm} \times \left(\pm \eta_{\pm} \mathbf{t} \pm\cos\frac{\phi}{2}\mathbf{n}- \sin\frac{\phi}{2}\mathbf{b}\right)  \\
&\quad\quad\quad\quad\quad\quad\quad\quad\quad\quad+\bm{\Omega}_{\pm} \times \left(\bm{\Omega}_{\pm} \times \left(\pm \eta_{\pm} \mathbf{t} \pm\cos\frac{\phi}{2}\mathbf{n}- \sin\frac{\phi}{2}\mathbf{b}\right)\right) \bigg) 
\\&\quad\quad\quad\quad\quad\quad\quad\quad\quad\quad\mp \frac{1}{4}v_{0\pm}^4 \dot{\eta}_{\pm} \left(\pm \eta_{\pm} \mathbf{t} \pm\cos\frac{\phi}{2}\mathbf{n}- \sin\frac{\phi}{2}\mathbf{b}\right) \times \left(\bm{\Omega}_{\pm}^{\prime} \times \left(\pm \eta_{\pm} \mathbf{t} \pm\cos\frac{\phi}{2}\mathbf{n}- \sin\frac{\phi}{2}\mathbf{b}\right)\right) \bigg], \\
&\mathbf{M}_{\pm} =  \frac{1}{\kappa_{n\pm}} \left(-\frac{\partial}{\partial S} \frac{\partial \varepsilon_{\pm}}{\partial \eta_{\pm}^{\prime}} + \frac{\partial \varepsilon_{\pm}}{\partial \eta_{\pm}}\right) \mathbf{t} + \left( m_3 \sin\frac{\phi}{2} -  \left(\frac{\partial \varepsilon_{\pm}}{\partial \kappa_{n\pm}} + \frac{\eta_\pm}{\kappa_{n\pm}}(\frac{\partial}{\partial S} \frac{\partial \varepsilon_{\pm}}{\partial \eta_{\pm}^{\prime}} - \frac{\partial \varepsilon_{\pm}}{\partial \eta_{\pm}})\right) \cos\frac{\phi}{2} \right) \mathbf{n}\\&\quad\quad \pm \left( m_3 \cos\frac{\phi}{2} + \left(\frac{\partial \varepsilon_{\pm}}{\partial \kappa_{n\pm}} + \frac{\eta_\pm}{\kappa_{n\pm}}(\frac{\partial}{\partial S} \frac{\partial \varepsilon_{\pm}}{\partial \eta_{\pm}^{\prime}} - \frac{\partial \varepsilon_{\pm}}{\partial \eta_{\pm}})\right) \sin\frac{\phi}{2} \right) \mathbf{b},\\
&\mathbf{M}_+\cdot\mathbf{t}+k_c(2\arctan(\frac{\kappa_n}{\kappa_g})-\phi_0)=\rho h \bigg[ \left(\frac{1}{2}v_{0+}^2 + \frac{1}{3}v_{0+}^3 \gamma_{+}\right) \left(+ \eta_{+} \mathbf{t} +\cos\frac{\phi}{2}\mathbf{n}- \sin\frac{\phi}{2}\mathbf{b}\right) \times \ddot{\mathbf{r}}_0  \\
&\quad\quad\quad\quad\quad\quad\quad\quad\quad\quad+\left(\frac{1}{3}v_{0+}^3 + \frac{1}{4}v_{0+}^4 \gamma_{+}\right) \left(+ \eta_{+} \mathbf{t} +\cos\frac{\phi}{2}\mathbf{n}- \sin\frac{\phi}{2}\mathbf{b}\right) \times \bigg( \dot{\bm{\Omega}}_{+} \times \left(+ \eta_{+} \mathbf{t} +\cos\frac{\phi}{2}\mathbf{n}- \sin\frac{\phi}{2}\mathbf{b}\right)  \\
&\quad\quad\quad\quad\quad\quad\quad\quad\quad\quad+\bm{\Omega}_{+} \times \left(\bm{\Omega}_{+} \times \left(+ \eta_{+} \mathbf{t} +\cos\frac{\phi}{2}\mathbf{n}- \sin\frac{\phi}{2}\mathbf{b}\right)\right) \bigg) 
\\&\quad\quad\quad\quad\quad\quad\quad\quad\quad\quad- \frac{1}{4}v_{0+}^4 \dot{\eta}_{+} \left( \eta_{+} \mathbf{t} +\cos\frac{\phi}{2}\mathbf{n}- \sin\frac{\phi}{2}\mathbf{b}\right) \times \left(\bm{\Omega}_{+}^{\prime} \times \left( \eta_{+} \mathbf{t} +\cos\frac{\phi}{2}\mathbf{n}- \sin\frac{\phi}{2}\mathbf{b}\right)\right) \bigg]. \\
\end{aligned}
\end{equation}
We then derive the component form of Eq. \eqref{B1} expressed in the fixed frame $xyz$. The correlations between $\mathbf{t},\mathbf{n},\mathbf{b}$ and $\mathbf{x},\mathbf{y},\mathbf{z}$ are expressed with Euler angles $\psi,\theta,\varphi$. The rotation transformation is defined as follows: 1. rotate about $\mathbf{z}$ through an angle $\psi$, and the frame $xyz$ transforms into $x_1y_1z$, 2. rotate about $\mathbf{x}_1$ through an angle $\theta$, and $x_1y_1z$ transforms into $x_1y_2b$, 3. rotate about $\mathbf{b}$ through an angle $\varphi$, and $x_1y_2b$ transforms into $tnb$.
Under this definition, the correlations between $\mathbf{t},\mathbf{n},\mathbf{b}$ and $\mathbf{x},\mathbf{y},\mathbf{z}$ are
\begin{equation}\label{B2}
\begin{pmatrix} \mathbf{t} \\ \mathbf{n} \\ \mathbf{b}\end{pmatrix} = \begin{pmatrix}
\cos\varphi \cos\psi - \cos\theta \sin\varphi \sin\psi & \sin\varphi \cos\psi + \cos\theta \cos\varphi \sin\psi & \sin\theta \sin\psi \\
-\cos\varphi \sin\psi - \cos\theta \sin\varphi \cos\psi & -\sin\varphi \sin\psi + \cos\theta \cos\varphi \cos\psi & \sin\theta \cos\psi \\
\sin\theta \sin\varphi & -\sin\theta \cos\varphi & \cos\theta
\end{pmatrix}\begin{pmatrix} \mathbf{x} \\ \mathbf{y} \\ \mathbf{z}\end{pmatrix}
\end{equation}
Define $\mathbf{\omega}_f,\mathbf{\Omega}_f$ by $(\mathbf{t}', \mathbf{n}', \mathbf{b}') = \mathbf{\omega}_f \times (\mathbf{t}, \mathbf{n}, \mathbf{b})$, $(\dot{\mathbf{t}}, \dot{\mathbf{n}}, \dot{\mathbf{b}}) = \mathbf{\Omega}_f \times (\mathbf{t}, \mathbf{n}, \mathbf{b})$. The variables could be expressed in terms of the Euler angles
\begin{equation}
    \begin{aligned}
\mathbf{\omega}_f&=(\psi'\sin\theta\sin\varphi+\theta'\cos\varphi)\mathbf{t}+(\psi'\sin\theta\cos\varphi-\theta'\sin\varphi)\mathbf{n}+(\psi'\cos\theta+\varphi')\mathbf{b},\\
\mathbf{\Omega}_f&=(\dot{\psi}\sin\theta\sin\varphi+\dot{\theta}\cos\varphi)\mathbf{t}+(\dot{\psi}\sin\theta\cos\varphi-\dot{\theta}\sin\varphi)\mathbf{n}+(\dot{\psi}\cos\theta+\dot{\varphi})\mathbf{b}.
    \end{aligned}
\end{equation}
With the definition of the Frenet frame, we then derive
\begin{equation}\label{eqeulerkt}
    \begin{aligned}
        &\kappa=\psi'\cos\theta+\varphi',\\
        &\tau=\psi'\sin\theta\sin\varphi+\theta'\cos\varphi,\\
        &\psi'\sin\theta\cos\varphi-\theta'\sin\varphi=0.
    \end{aligned}
\end{equation}

Since $\mathbf{r}_0=x\mathbf{x}+y\mathbf{y}+z\mathbf{z}$ and $\mathbf{r}'_0=\mathbf{t}$, the geometric correlations exist between the Euler angles and the position coordinates by
\begin{equation} \label{xyz1}
    \begin{aligned}
        x'=\cos\varphi \cos\psi - \cos\theta \sin\varphi \sin\psi,\quad y'=\sin\varphi \cos\psi + \cos\theta \cos\varphi \sin\psi, \quad z'= \sin\theta \sin\psi.
    \end{aligned}
\end{equation}
Substituting Eqs. \eqref{framedtof},\eqref{avelo} \eqref{B2}-\eqref{eqeulerkt} into \eqref{B1} yields 7 equations related to $x,y,z,\psi,\theta,\varphi,F_{x\Sigma},F_{y\Sigma},F_{z\Sigma},m_{3\pm}$. Along with Eq. \eqref{xyz1} and $\psi'\sin\theta\cos\varphi-\theta'\sin\varphi=0$, we derive 11 equations with 11 variables, which can be solved numerically in a closed form.   

For a single developable surface, the equations and the process are similar. The specific form of the dynamic equations is similar to Eq. \eqref{B1} by removing the summation and the negative sign, and the fourth equation in Eq. \eqref{B1} no longer exists. Then the variables are $x,y,z,\psi,\theta,\varphi,F_{x},F_{y},F_{z},m_{3}$, and there are 10 equations with 10 variables, which can thus be solved numerically. The process is illustrated in the flowchart below.
\begin{figure}[!h]
    \centering
    \includegraphics[width=1.0\textwidth]{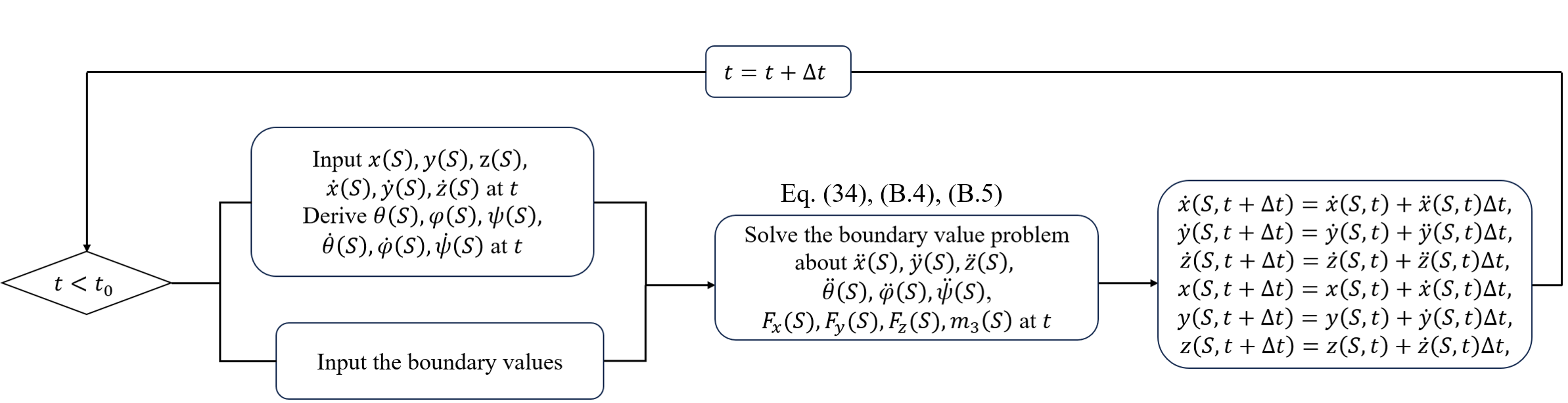}
\end{figure}
\bibliographystyle{elsarticle-harv} 
\bibliography{cas-refs}





\end{document}